\documentclass[fleqn,usenatbib]{mnras}

\usepackage{newtxtext,newtxmath}
\usepackage[T1]{fontenc}

\DeclareRobustCommand{\VAN}[3]{#2}
\let\VANthebibliography\thebibliography
\def\thebibliography{\DeclareRobustCommand{\VAN}[3]{##3}\VANthebibliography}

\usepackage{graphicx}	% Including figure files
\usepackage{amsmath}	% Advanced maths commands

\usepackage{amssymb}	% Extra maths symbols
\usepackage{url}
\title[Stellar population gradients of nine infalling dEs]{On the accretion of a new group of galaxies onto Virgo: III. The stellar population radial gradients of dEs}

\author[B. Bidaran et al.]{
Bahar Bidaran,$^{1,2}$\thanks{E-mail: Bidaran@ugr.es (BB)}
Francesco La Barbera,$^{3}$
Anna Pasquali,$^{2}$
Glenn van de Ven,$^{4}$ 
Reynier Peletier,$^{5}$
\newauthor
Jesus Falc\'on-Barroso,$^{6,7}$
Dimitri A. Gadotti,$^{8}$
Agnieszka Sybilska,$^{9}$
Eva K. Grebel$^{2}$
\\
$^{1}$Dpto. de F\'\i sica y del Cosmos, Campus de Fuentenueva, Edificio Mecenas, Universidad de Granada, E18071--Granada, Spain\\
$^{2}$Astronomisches Rechen-Institut, Zentrum f\"ur Astronomie der Universit\"at Heidelberg, M\"onchhofstra\ss e 12-14, 69120 Heidelberg, Germany\\
$^{3}$INAF-Observatorio Astronomico di Capodimonte, sal. Moiariello 16, Napoli, 80131, Italy\\
$^{4}$Department of Astrophysics, University of Vienna, T\"urkenschanzstra{\ss}e 17, 1180 Vienna, Austria \\
$^{5}$Kapteyn Astronomical Institute, University of Groningen, Postbus 800, 9700 AV Groningen, the Netherlands\\
$^{6}$Instituto de Astrof\'isica de Canarias, Calle V\'ia L\'actea s/n, E-38205 La Laguna, Tenerife, Spain\\
$^{7}$Departamento de Astrof\'isica, Universidad de La Laguna, E-38200 La Laguna, Tenerife, Spain \\
$^{8}$Centre for Extragalactic Astronomy, Department of Physics, Durham University, South Road, Durham DH1 3LE, UK\\
$^{9}$Sybilla Technologies, Torunska 59, 85-023 Bydgoszcz, Poland\\
}

\date{Accepted XXX. Received YYY; in original form ZZZ}

\pubyear{2020}

\begin{document}
\label{firstpage}
\pagerange{\pageref{firstpage}--\pageref{lastpage}}
\maketitle

\begin{abstract}
Using MUSE data, we investigate the radial gradients of stellar population properties (namely age, [M/H], and the abundance ratio of $\alpha$ elements [$\alpha$/Fe]) for a sample of nine dwarf early-type (dE) galaxies with log(M$_{\star}$/M$_{\odot}$) $\sim$ 9.0 and an infall time onto the Virgo cluster of 2-3Gyr ago. We followed a similar approach as in Bidaran et al. (2022) to derive their stellar population properties and star formation histories (SFHs) through fitting observed spectral indices and full spectral fitting, respectively. We find that these nine dE galaxies have truncated [Mg/Fe]vs.[Fe/H] profiles than equally-massive Virgo dE galaxies with longer past infall times. Short profiles of three dE galaxies are the result of their intense star formation which has been quenched long before their accretion onto the Virgo cluster, possibly as a result of their group environment. In the remaining six dE galaxies, profiles mainly trace a recent episode of star burst within 0.4R$_{\rm e}$ which results in higher light-weighted [$\alpha$/Fe] values. The latter SFH peak can be due to ram pressure exerted by the Virgo cluster at the time of the accretion of the dE galaxies. Also, we show that younger, more metal-rich and less $\alpha$-enhanced stellar populations dominate their inner regions (i.e., < 0.4R$_{\rm e}$) resulting in mainly flat  $\nabla_{\rm age}$, negative $\nabla_{\rm [M/H]}$ and positive  $\nabla_{\rm [\alpha/Fe]}$. We find that with increasing log($\sigma_{\rm Re}$) of dE galaxies, $\nabla_{\rm age}$ and  $\nabla_{\rm [\alpha/Fe]}$ flatten, and the latter correlation persists even after including early-type galaxies up to log($\sigma_{\rm Re}$ $\sim$ 2.5), possibly due to the more extended star formation activity in the inner regions of dEs, as opposed to more massive early-type galaxies. 
\end{abstract}


%%%%%%%%%%%%%%%%%%%%%%%%%%%%%%%%%%%%%%%%%%%%%%%%%%

%%%%%%%%%%%%%%%%% BODY OF PAPER %%%%%%%%%%%%%%%%%%

\section{Introduction}\label{Introduction}

Early-type dwarf galaxies (dEs) with log(M$_{\star}$ [M$_{\odot}$]) < 9.5 constitute a large yet diverse population in the Local Universe, particularly in galaxy groups and clusters \citep{1974Oemler, 1980Dressler, 1993Whitmore, 1984Postman,2004Kauffmann,2010Peng,2019Davies}. Dwarf ellipticals are generally pressure-supported systems with low rates of star formation, if any \citep[e.g.,][]{2011Toloba, 2015MNRAS.452.1888R, 2020Scott}. Detailed studies revealed diversities in their stellar population properties, kinematics, and morphologies, based on which various formation scenarios have been proposed. Some propose dEs as the remnants of low-mass late-type galaxies that were transformed by their environment \citep{1985Kormendy,1988Binggeli,2014Boselli}. In other scenarios dEs are primordial objects, formed in the early Universe and evolved passively (i.e., did not go through any mergers) \citep[e.g.,][]{2003DeRijcke, 2017Wheeler}. To decode formation and evolution of dEs, one way is to investigate in detail their stellar population properties, as they retain the fossil record of their mass assembly and evolution histories. 

Primordial galaxies are presumed to be formed through dissipative (monolithic) collapse \citep{1974Larson,1987Arimoto,2009Dekel} of gas clouds which grows more metal rich as the result of galaxy's own stellar evolution. As the galaxy ages further gas sinks toward the galaxy's center powering the formation of new stars in the inner regions. Within this context, the orbits within which stars form remain untouched, thus, a positive stellar age and a rather steep negative stellar metallicity gradients are expected across the galaxy. This formation scenario explains the observed radial gradients of [M/H], [$\alpha$/Fe], reddening of colours \citep[e.g.,][]{1990Peletier,2022Liao} and line strength indices \citep[e.g.,][]{1993Carollo,1993Davies,2000Trager} in ETGs. Observations of \cite{1999Kobayashi}, in agreement with predictions from hydrodynamical simulations of \cite{2008Pipino}, show a slope range of -0.5 to -0.2, per decade of radius, for the stellar metallicity gradient ($\nabla_{\rm [M/H]}$) in early-type galaxies (ETGs) over a wide range of stellar mass. Investigations of the inner regions of massive early-/late-type galaxies \citep[e.g.,][]{2005Forbes,2011LaBarbera,2017Goddard,2018MNRAS.475.3700M,2019Ferreras,2020Zibetti} also endorse the presence of such a negative stellar metallicity gradient. Moreover, the radial gradients of stellar populations in cluster dEs, and similarly some of the Local Group dwarfs \citep[e.g.,][]{2015Ho}, support a scenario where star formation has been prolonged and mostly concentrated in the galaxy center \citep[see also][]{2010Paudel}. 

The observational evidence for how environment can possibly modify radial distribution of stellar population properties in ETGs is still controversial. For instance, in the stellar mass range of 10$^{9.5-12}$ [M$_{\odot}$], \cite{2020Santucci} show that satellite and central passive galaxies exhibit similar stellar population gradients, within 2R$_{\rm e}$, independent of their stellar mass or halo size \citep[see also][]{2017Zheng}. On the other hand, \cite{2011LaBarbera} showed that ETGs in galaxy groups tend to show steeper stellar age and metallicity gradients, within 1R$_{\rm e}$, compared to their field counterparts, indicating a possible correlation between galaxies environment and their stellar populations radial gradients \citep[see also][]{2019Ferreras}. 
Intriguingly, the  stellar population radial gradients in low-mass galaxies (i.e., M$_{\star}$ $<$ 10$^{9.5}$ [M$_{\odot}$]) have been seldomly investigated thus far in the literature \citep[relevant studies are ][]{2008Chilingarian,2009Koleva,2011Koleva,2017Goddard,2017Sybilska}.

\setlength{\tabcolsep}{3.pt}
\begin{table*}
\caption{\label{BL_sample} Properties of the main sample}
\centering
\begin{tabular}{c c l l c c c c c c c}
\hline
Object      & type    & $\alpha$ (J2000)    &   $\delta$ (J2000)  &  z$^{a}$  & \textit{R}$_{\rm e}^{a}$ \rm [arcsec] & \textit{M}$_{\rm r}^{a}$ \rm [mag] & \textit{g-r}$^{a}$ \rm [mag] &  $ A_{\rm V}^{b}$ [mag]   & $\sigma_{\rm Re}$ $^{\rm c}$ [kms$^{-1}$] & T$_{\rm exp}$ [hour]\\
\hline
\hline
VCC 0170 & dE(bc) & 12 15 56.30 & +14\ 25\ 59.2 & 0.00472 & $31\farcs57$ & -17.62 & 0.59  & 0.089& 27.1 $\pm$ 10.6 & 4\\
VCC 0407 & dE(di) & 12 20 18.80 & +09\ 32\ 43.1 & 0.00626 & $18\farcs38$ & -17.37 & 0.61  & 0.057 & 32.3 $\pm$ 8.8 &2\\
VCC 0608 & dE(di) & 12 23 01.70 & +15\ 54\ 20.2 & 0.00607 & $25\farcs77$ & -17.58 & 0.60  & 0.072 & 25.1 $\pm$ 9.2 &5\\
VCC 0794 & dE(nN) & 12 25 21.61 & +16\ 25\ 46.9 & 0.00558 & $37\farcs33$ & -17.29 & 0.61  & 0.065 & 33.0 $\pm$ 7.5 &3\\
VCC 0990 & dE(di) & 12 27 16.93 & +16\ 01\ 28.1 & 0.00573 & $10\farcs31$ & -17.43 & 0.62  & 0.080 & 36.0 $\pm$ 5.6 &3\\
VCC 1833 & ---    & 12 40 19.70 & +15\ 56\ 07.1 & 0.00569 & $8\farcs52$  & -17.44 & 0.61  & 0.099 & 34.4 $\pm$ 6.0 & 3\\
VCC 1836 & dE(di) & 12 40 19.50 & +14\ 42\ 54.0 & 0.00668 & $42\farcs27$ & -17.45 & 0.58  & 0.079 & 38.5 $\pm$ 8.2 & 4\\ 
VCC 1896 & dE(di) & 12 41 54.60 & +09\ 35\ 04.9 & 0.00629 & $14\farcs98$ & -17.04 & 0.62  & 0.047 & 27.0 $\pm$ 7.2 & 3\\
VCC 2019 & dE(di) & 12 45 20.40 & +13\ 41\ 34.1 & 0.00607 & $18\farcs60$ & -17.65 & 0.63  & 0.060 & 31.2 $\pm$ 6.5 & 2\\
\hline
\end{tabular}\\
Columns are: Name of target, morphological type, RA and DEC, redshift (z), effective radius (R$_{\rm e}$) (half-light major axis), absolute r-band magnitude (M$_{r}$), \textit{g-r} colour measured at R$_{\rm e}$, foreground Galactic extinction in V-band (A$_{V}$), stellar velocity dispersion at R$_{\rm e}$ ($\sigma_{\rm Re}$), total exposure time (TET). \\
a:\,\cite{2006Lisker},
b:\,\cite{2011Schlafly},
c:\,\cite{2020Bidaran}
\\
\end{table*}

Using long-slit spectroscopy, \cite{2011Koleva} (hereafter K11) investigated 40 galaxies within the mass range of 10$^{7-12}$ [M$_{\odot}$], consisting of 14 dEs/dS0s plus lenticulars and elliptical galaxies in the Fornax and Virgo clusters. They measured negative metallicity gradients and moderately positive age gradients for Fornax dEs. Except for faint dEs, they could not detect any meaningful correlation between age or metallicity gradients of dEs and their stellar velocity dispersion, stellar mass, or luminosity. Based on these results, K11 concluded that the stellar population radial gradients in dwarf galaxies reflect those of their progenitors, and their environment had minimal effect on altering them. Similarly, using a sample of 20 Virgo dEs and 258 massive early-type galaxies from the ATLAS3D survey \citep{2011Cappellari},  \cite{2017Sybilska} (hereafter S17) could not detect any meaningful correlation between stellar population radial gradients and the stellar velocity dispersion of early-type galaxies. Unfortunately, these studies cover only a limited number of dEs (less than 50 dEs in total).

In \cite{2022Bidaran} (hereafter B22), we analysed the stellar population properties of a sample of nine dEs in the Virgo cluster that, based on their distribution on the projected phase-space diagram and predictions of N-body simulations \citep{2015Vijayaraghavan}, are expected to have been accreted onto Virgo about 2-3 Gyr ago \citep{Lisker2018}. According to the N-body simulations, these nine dEs were likely part of a massive galaxy group with M$_{\star}$ $\sim$ 10$^{13}$ [M$_{\odot}$], prior to their merger with Virgo. In B22 we investigated their integrated stellar population properties using their averaged MUSE (Multi-Unit Spectroscopic Explorer) spectra. We showed that they are, on average, metal-poor and noticeably $\alpha$-enhanced than other equally-massive dEs in the Coma and Virgo clusters with similar or even larger accretion time. We detected prominent peaks in the star formation history of six dEs in this sample, indicating an enhanced star formation activity during or after their accretion onto Virgo, possibly due to ram pressure. We also showed that the other three dEs had been already quenched, possibly in a group environment, before their accretion onto Virgo. In this study, we proceed further and present a detailed analysis of spatially resolved stellar population properties of these nine dEs using their MUSE data. In particular, we are interested in whether their stellar population radial gradients are affected by their (present and past) environment and intrinsic properties (e.g., stellar mass). For this purpose, we compare our results with available analyses of dEs in the Virgo and Fornax clusters and more massive ETGs from the literature.  

This paper is organized as follows: In Section \ref{Data} we briefly introduce the main sample of this work as well as the other comparison samples employed in the analysis. In Section \ref{Analysis} we describe our methods for deriving the stellar population properties  and star formation histories of the main sample. In Section \ref{Results} we present the stellar population maps of our main sample and quantify the spatial distribution of their stellar population properties by computing stellar population radial gradients. We also present their star formation histories as computed within and outside 0.4R$_{\rm e}$. Furthermore, we compare our results with the literature. We discuss our results in Section \ref{Discussion} and summarize them in Section \ref{Conclusion}.

%--------------------------------------------------------------------------------
\section{Data}\label{Data}
\subsection{The main sample}\label{Data1}
The main goal of this paper is to analyse the spatial distribution of the stellar population properties of nine Virgo dEs within the magnitude range of -17$\geq$M$_{r}$ [mag]>-18\footnote{within the stellar mass range of 8.9 < log (M$_\star$ [M$_\odot$]) < 9.2, as estimated in B20.} which fell into Virgo 2-3 Gyr ago. Their selection criteria and MUSE data reduction are extensively described in \cite{2020Bidaran} (hereafter B20). In short, we observed these nine Virgo dEs using the MUSE instrument mounted on the Very Large Telescope (VLT). MUSE covers the wavelength range of 4750 to 9350 \AA\ with a wavelength-dependent spectral resolution. The average spectral resolution is 2.51 \AA\ (full width at half maximum; FWHM) \citep{2010Bacon}. MUSE has a spatial resolution of 0.2 arcsec/pixel and covers a field of view of 1$\times$1 arcmin$^{2}$. Observations were acquired with a nearly constant seeing ($\sim$ 1.6'') as part of a science verification proposal, in the period December 2016 to February 2017 and February 2018 to July 2018 (P98, ESO programs 098.B-0619 and 0100.B-0573; PI: Lisker). 

In Table \ref{BL_sample} we summarise the main properties of our main sample, including the ID of each galaxy, morphological type, right ascension and declination, redshift, effective radius (R$_{\rm e}$) in the r-band, absolute r-band magnitude, g-r color, foreground Galactic extinction in V-band (A$_{V}$), stellar velocity dispersion at 1R$_{\rm e}$, and total exposure time (T$_{\rm exp}$) of the MUSE observations used in this study. The A$_{V}$ values are taken from \cite{2011Schlafly} using the NASA/IPAC Extragalactic Database (NED). The M$_{\rm r}$ and g-r colors reported in Table \ref{BL_sample} are from \cite{2006Lisker} and \cite{2008Janz,2009Janz} and are both corrected for Galactic extinction. The $\sigma_{\rm Re}$ values come from B20. The remaining properties in Table \ref{BL_sample} are taken from \cite{2006Lisker}.

\subsection{The Comparison samples}\label{Data2}
Throughout this work, we compare properties of our main sample with information available in the literature for dEs in the Virgo and Fornax clusters and more massive ETGs. Our comparison sample includes early-type galaxies from the following studies: 

\begin{itemize}
    \item \textbf{Virgo dEs:} We use 17 Virgo dEs within the stellar mass range of 9 $<$ log(M$_{\star}$ [M$_{\odot}$]) $<$ 9.7 that were observed with the SAURON spectrograph at the William Herschel Telescope. Their stellar population properties were estimated by \cite{2017Sybilska} from absorption features (H$\beta_{0}$, H$\beta$, Fe4383, Fe5015, Mgb), using the MILES  stellar population models. 

    \item \textbf{Fornax dEs: } We use 13 Fornax dEs within the magnitude range of -17.9 < M$_{B}$ [mag] < -15.7 that were observed with the FORS2 spectrograph on the VLT and Gemini Multi Object Spectrographs (GMOS). Their stellar population properties were derived by \cite{2011Koleva} through full spectrum fitting, using the Pegase.HR 4 stellar population models \citep{2004LeBorgne}. 

    \item \textbf{ETGs from the MANGA survey}: \cite{2021Parikh} covers 1900 galaxies in a wide stellar mass range of 8.6 $<$ log(M$_{\star}$ [M$_{\odot}$]) $<$ 11.3. Their stellar population properties were measured by modelling absorption features (H$\beta$, Mgb, Fe5070, Fe5335) using the TMJ stellar population models. It should be noted that \cite{2021Parikh} binned galaxies spectra according to stellar mass, and measured spectral indices over the averaged spectrum of all galaxies within each stellar mass bin. Here, we compare our results with those from \cite{2021Parikh}) for six stellar mass bins, each including $\sim$ 150 galaxies. 
    
    \item \textbf{ETGs from the ATLAS3D survey}: 260 ETGs were observed with the SAURON IFU instrument and cover a stellar mass range of 10 $<$ log(M$_{\star}$ [M$_{\odot}$]) $<$ 12. Their stellar population properties were obtained by \cite{2020A&A...635A.129K}, from absorption features (H$\beta$, Mgb, and Fe5015) using \cite{2007Schiavon} single stellar population (SSP) models.
    
    \item \textbf{Massive ETGs from the CALIFA survey}: We use 45 massive ETGs within the stellar mass range of 10.4 $<$ log(M$_{\star}$ [M$_{\odot}$]) $<$ 11.5 that were observed with the PMAS/PPAK IFU instrument. Their stellar population properties were derived by \cite{2018MNRAS.475.3700M} by modelling absorption features (H$\beta_{0}$, Fe4383, Fe5015, Fe5270, Mgb) with the MILES stellar population models. 
\end{itemize}

\subsection{Single stellar population models}\label{SSP_models}
Our analysis relies on \cite{2010Vazdekis,2015Vazdekis} single stellar population (SSP) models based on the MILES stellar library \citep{2006MNRAS.371..703S,2007Cenarro,2011A&A...532A..95F}. These SSP models, with an original spectral resolution of FWHM = 2.51 \AA\, are constructed based on the BASTI isochrones \citep{2004Pietrinferni} and a bi-modal initial mass function (IMF) with a slope of 1.3 \citep{1996Vazdekis}, which is very similar to a Kroupa IMF. 

Given our results in B22, we choose SSP models covering the age range of 0.5 to 14 Gyr, the [M/H] range of –1.26 to -0.25 dex, and two [$\alpha$/Fe] values of 0.00 dex and 0.40 dex. To obtain more precise fits we linearly interpolate the SSP models in age-[M/H]-[$\alpha$/Fe]-[index] space, using steps of 0.02 dex and 0.015 dex in [M/H] and [$\alpha$/Fe], respectively. To maintain an approximately constant variation of age-sensitive spectral indices (i.e., Balmer lines) with age, we interpolate young and old models using different steps of 0.01 and 0.04 Gyr, respectively. Furthermore,  we extrapolate the models with a constant step of 0.015 dex, to cover a final [$\alpha$/Fe] range of -0.3 dex to 0.7 dex. This range was selected following the work of \cite{2009Smith} and \cite{2017Sybilska}, who show that dEs in the Coma and Virgo clusters span a typical range of -0.3 $<$ [$\alpha$/Fe] [dex] $<$ 0.6. 

%-----------------------------------------------------------------------------
\section{Analysis}\label{Analysis}
\subsection{Fitting the Spectral indices}\label{Analysis_1}
First, we mask foreground stars and background galaxies in the MUSE data cube of each dE. Afterward, we discard spaxels with a signal-to-noise ratio (SNR)$<$3 and spatially bin the MUSE data using the adaptive spatial Voronoi binning method, developed by \cite{2003Cappellari}. This method is a binning algorithm that assigns spaxels to different bins in such a way that the ultimate SNR of each bin meets a defined minimum threshold. Given the size of the dEs in our sample and their surface brightness range, we find a minimum SNR of 100 per bin to be an appropriate choice for measuring stellar population properties. SNR $\geq$ 100 provides an optimal compromise between the accuracy of the derived stellar population properties and the spatial sampling of the dEs outskirts \citep[see e.g.,][]{2020Bittner}. Each binned spectrum is the average of the spectra of all spaxels within the bin.

We correct all the binned spectra for Galactic foreground extinction using the A$_{V}$ values as reported in Table \ref{BL_sample} and adopting R$_{v}$ = 3.10, following the Cardelli Galactic extinction law \citep{1989Cardelli}. We then correct the binned spectra for possible tracers of nebular emission lines using the GANDALF package \citep[Gas AND Absorption Line Fitting;][]{2006Sarzi, 2006MNRAS.369..529F}. This algorithm uses a Gaussian function to simultaneously fit emission lines in the observed spectra. We subtract the emission line spectra measured by GANDALF from the observed binned ones, and later, we use these corrected binned spectra to measure line indices and derive the stellar population properties.

Our fitting procedure derives stellar population properties by comparing observed spectral indices that are age- ( H$\beta_{0}$, see Table \ref{Index banpass}) and metallicity-sensitive (Mg, Fe5070, and Fe5335) with predictions of SSP models. The Lick indices employed in this work (Table \ref{Index banpass}) are defined in the LIS-5 \AA\ system \citep[FWHM $=$ 5\AA\.][]{1996Vazdekis}, and to achieve that resolution, we convolve our binned spectra with a Gaussian kernel whose FWHM$_{\rm final}$ = $\sqrt{\rm FWHM_{\rm LIS5.00}^{2} - (\rm FWHM_{\rm MUSE}^{2} + \rm FWHM_{\rm \sigma}^{2})}$. Here, FWHM$_{\rm LIS5.00}$ is the resolution of the LIS-5 \AA\ system and FWHM$_{\rm MUSE}$ is the wavelength-dependent MUSE resolution \citep{2010Bacon}. FWHM$_{\rm \sigma}$ corresponds to the stellar velocity dispersion of the binned spectrum as measured by B20, using the pPXF \citep[][]{2004Cappellari, 2017Cappellari}. We also broaden the SSP models from their original spectral resolution to the LIS$-$5 \AA\ system. As in B22, we consider three uncertainty sources to estimate the random error on each measured index: the redshift estimate, the spectral flux, and the stellar velocity dispersion \citep{2006Kuntschner}. We perform 125 Monte Carlo (MC) iterations for each index, each time perturbing the observed flux based on the three mentioned uncertainty sources. The standard deviation of these multiple measurements is assumed as the index error. To avoid unrealistically small errors on stellar population parameters, as thoroughly explained in B22, we add a minimum systematic uncertainty of 0.05 \AA\ in quadrature to the errors estimated for each index.  

The stellar population properties of our targeted dEs (namely their age, [M/H] and [$\alpha$/Fe]) are derived using an iterative fitting procedure, based on the $\chi^2$ minimization approach, developed in B22. A comprehensive explanation of this method is presented in B22, but for the reader’s convenience, we briefly summarize it here.
Our fitting approach consists of four steps and, in each step, we fit specific pairs of indices: (1) We first perform 100 fits over the finer grid of H$\beta_{0}$ vs. MgFe, constructed for models with [$\alpha$/Fe]= 0.0 dex. This fit gives us an initial estimate of the dE’s age. (2) We then construct a grid of Mgb-<Fe> for all SSP models within the age range retrieved from the first step, and perform 100 fits to derive [$\alpha$/Fe]. (3) The [$\alpha$/Fe] range measured in this step is then used to construct a more precise H$\beta_{0}$ vs. MgFe grid and to measure age and [M/H]. We repeat step three for 100 random [$\alpha$/Fe] values taken from the results of step two, and run 200 MC realizations.
The final age and [M/H], reported per bin, are the mean values of 200 MC realizations in this step. (4) In the fourth step, we take 100 random values from the ages derived in step three, and perform 200 MC realizations to measure [$\alpha$/Fe]. The final [$\alpha$/Fe] values are the averages of the measurements in step four. By performing this complete loop of four fits, we measure the light-weighted age, [M/H], and [$\alpha$/Fe] properties of each binned spectrum. We assume the standard deviation of the values measured in steps three and four as their corresponding errors. We discuss results of these measurements in Section \ref{Results_1}.

\begin{figure*}
\includegraphics[scale=0.8]{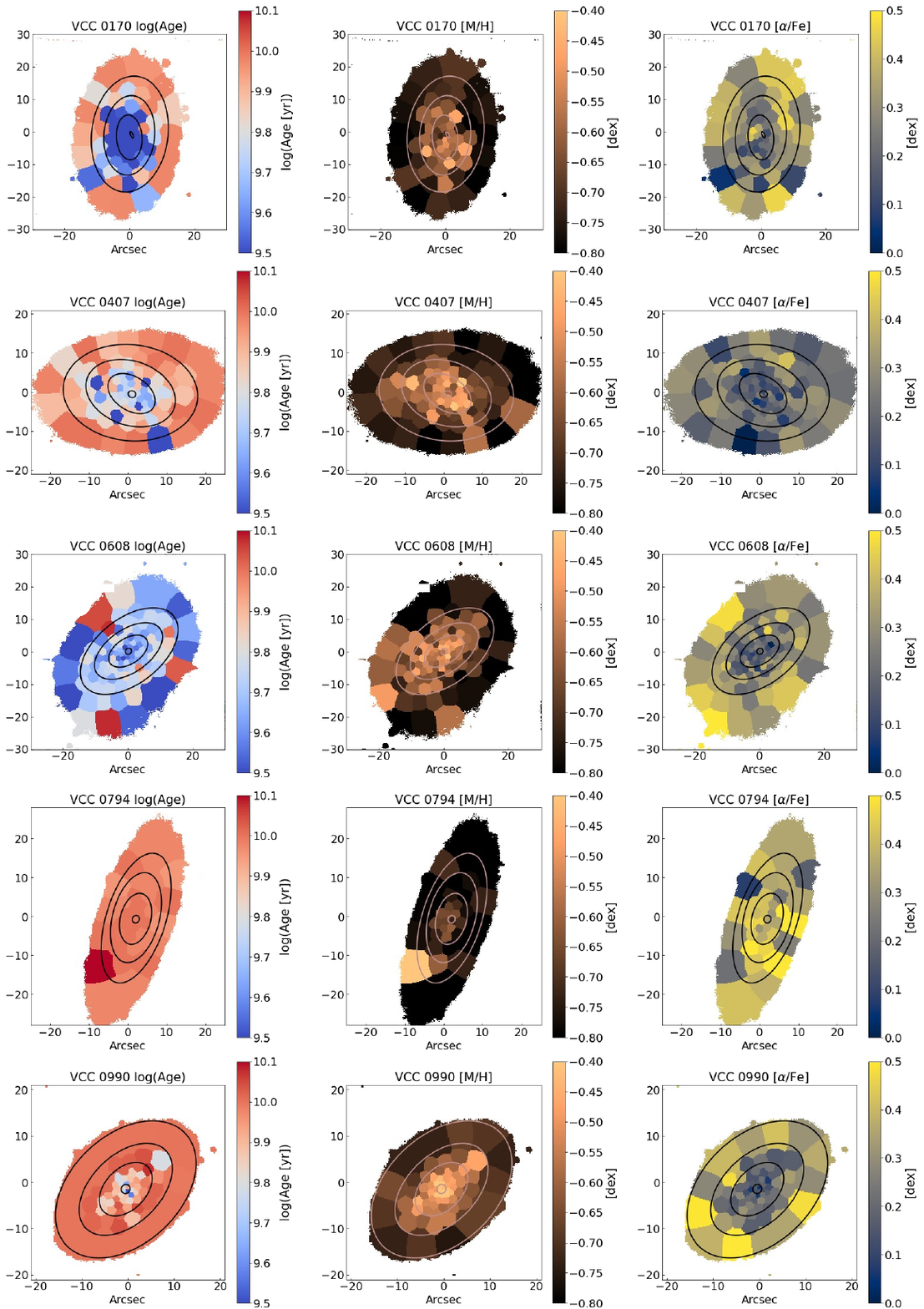}
\caption{Stellar population maps of the main sample dEs. For each dE and from left to right, Age, [M/H] and [$\alpha$/Fe] maps are presented. On all three maps, isophotes (measured from MUSE stack images in B20) are overplotted. From the inside out, the isophotes indicate regions with a surface brightness of 20.59, 21.00, 21.85, and 22.52 ABmag/arcsec$^2$, respectively.}
\label{stellar pops maps1}
\end{figure*}

\begin{figure*}
\includegraphics[scale=0.8]{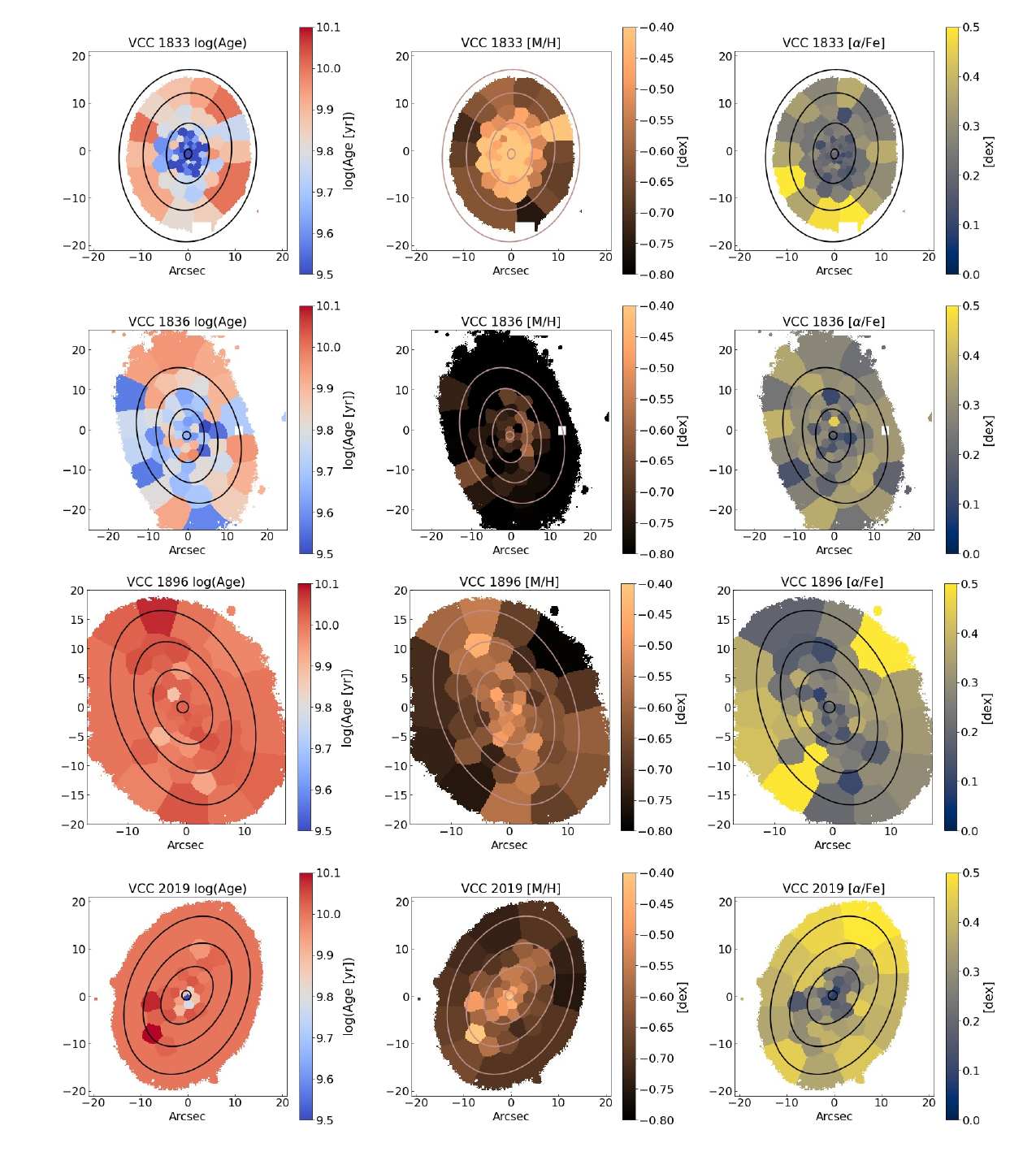}
\caption{Continued.}
\label{stellar pops maps2}
\end{figure*}

\subsection{Full spectrum fitting}\label{Analysis_2}
To have a complementary picture of the radial distribution of stellar population properties in dEs and an estimate of the star formation histories (SFHs) in their inner and outer regions, we utilize the full-spectrum fitting routine STARLIGHT. This routine linearly combines a set of SSP models to construct a synthetic spectrum that best represents the observed light \citep{2005CidFernandes}. We define an aperture with a radius 0.4R$_{\rm e}$ and calculate the sum of rest frame flux (i.e., redshift corrected), per wavelength, of all the spaxels within this aperture, to construct the inner integrated spectrum. Spectra of spaxels outside this aperture are also summed together to construct the integrated outer spectrum. Note that only pixels with SNR $>$ 3 are summed up. As explained in B22, two dEs of our sample, namely VCC0990 and VCC2019, host a central nuclear stellar cluster (NSC). For these two dEs, the inner spectra are constructed after masking the radius of the NSC (R$_{\rm NSC}$), as reported in B22. We find the radius of 0.4R$_{\rm e}$ to provide a better compromise for the spatial sampling of dEs’ outskirts, as the MUSE field of view is too small to cover dEs as extended as VCC1836 and VCC0794. Also, an aperture smaller than 0.4R$_{\rm e}$ would cause outer spectra to be noticeably over-weighted by the light of dEs’ inner parts.

We follow a similar approach as in B22 for the fitting. To briefly summarize it here, we perform the fitting within the spectral range of 475 to 554 nm. We use a similar set of SSP models as in Section \ref{Analysis_1} (meaning with the same isochrones and IMF) within the age range of 1 to 14 Gyr (with steps of 1 Gyr) and the metallicity range of -1.26 to -0.25 dex. For each galaxy we run STARLIGHT using three "SSP families", each of them consisting of models with fixed [$\alpha$/Fe] and varying age and [M/H]. For every galaxy, we construct three SSP families by interpolating and/or extrapolating original models based on the mean [$\alpha$/Fe] and the two $<$[$\alpha$/Fe]$>$ $\pm$ 1$\sigma$ values, measured from the integrated light of the galaxy in B22 (see Table 2 of that study). Errors of these fits are computed based on 100 MC iterations, as explained in B22. We discuss results of these measurements in Section \ref{Results_2}

%--------------------------------------------------------------------------
\section{Results}\label{Results}
\subsection{Age, Metallicity and [$\alpha$/Fe] maps}\label{Results_1}
We construct maps of age, [M/H] and [$\alpha$/Fe] for each galaxy, using results of spectral indices measurements from Section \ref{Analysis_1}. In the left-hand side column of Figs. \ref{stellar pops maps1} and \ref{stellar pops maps2}, we present the age maps of our main sample dEs. In most cases, the age maps are flat and show little variations over the galaxies. In the middle column of the same figures, we present the [M/H] maps which indicate the presence of more metal-rich stellar populations in the central regions of all nine dEs. In the right-hand side column of Figs. \ref{stellar pops maps1} and \ref{stellar pops maps2} we present the [$\alpha$/Fe] maps which show that most dEs are dominated by [$\alpha$/Fe]-enhanced stellar populations in their outskirts, with the [$\alpha$/Fe] decreasing toward the inner regions. In what follows, we describe some features specific to each dE:
\begin{itemize}
    \item \textbf{VCC0170:} As explained in B20, this dE is the only member of our main sample for which we detected nebular emission lines in its central regions. In B22 we showed that this dE experienced a peak in its star formation history at the time of its accretion onto Virgo ($\sim$ 2-3 Gyr ago), possibly triggered by ram pressure acting during the early phases of accretion. Ram pressure can compress gas in the galaxy’s inner disk and initiate or enhance an episode of star formation, mainly in the inner regions of infalling galaxies \cite[e.g., ][]{2021Boselli,1998Fujita,1999Fujita,2001Vollmer}. The presence of young and more metal-rich stellar populations in the central regions of VCC0170 and the fact that this dE is still actively forming new stars in its center (see B20), supports the picture presented above. 

    \item \textbf{VCC0407:} The stellar populations confined in the central regions of this dE are more metal-rich and somewhat younger than in its outskirts. We did not detect any variation in [$\alpha$/Fe] values across this galaxy. 

    \item \textbf{VCC0608:}  VCC0608 is also dominated by more metal-rich and [$\alpha$/Fe]-enhanced stellar populations in its inner regions. However, its age map is rather flat, showing no significant fluctuation in age values across the galaxy. 

    \item \textbf{VCC0794:} Our analysis in B22 shows that VCC0794 is the most metal-poor and $\alpha$-enhanced galaxy of our sample (See Table 2 of that study). The homogeneous distribution of old, metal-poor, and $\alpha$-enhanced stellar populations across this dE confirms our results in B22. 
    
    \item \textbf{VCC0990 and VCC2019:} Both dEs host a central NSC that shows distinct stellar population properties from the rest of their host galaxies \citep[e.g.,][]{2011Paudel,2021Fahrion}. As evident from the maps of these two dEs, their central NSC is, on average, younger, more metal-rich, and less $\alpha$-enhanced than the rest of the galaxy. The central NSC aside, these two dEs are dominated by old stellar populations, and exhibit similar radial variations of [M/H] and [$\alpha$/Fe], with a metal-poor and $\alpha$-enhanced population in the outer regions. 
    
    \item \textbf{VCC1833:} This galaxy is dominated by young, metal-rich, and less $\alpha$-enhanced stellar populations in its inner regions, which grow older and metal-poor at larger galactocentric distances. We do not see a clear variation in [$\alpha$/Fe] values across this galaxy. 
    
    \item \textbf{VCC1836:} This dE, with R$_{\rm e}$ = 6.05 kpc, is the most extended member of our sample and MUSE could only sample it up to $\sim$ 0.4 R$_{\rm e}$. Hence, the maps presented in Fig. \ref{stellar pops maps2} only show the stellar population properties of its inner regions. VCC1836 is dominated by stellar populations within the age range of 3 to 9 Gyr, the [M/H] range of -0.9 to -0.6 dex, and an average [$\alpha$/Fe] = 0.30 dex.

    \item \textbf{VCC1896:} The [M/H] and [$\alpha$/Fe] maps of this dE show the presence of more metal-rich and less $\alpha$-enhanced stellar populations along VCC1896’s photometric major axis. Meanwhile, the age map of this dE shows no particular variations as the old stellar populations dominate the entire galaxy. Deep photometric observations of VCC1896 \citep{2006Lisker,2009Lisker,2021Michea} proved the presence of a central bar and two faint spiral arms. Here, given the depth and spatial coverage of our MUSE data, we can only map the bright bar of VCC1896, which is extends over the VCC1896's photometric major axis. The distribution of [M/H] and [$\alpha$/Fe] resembles a V-shape profile (with the maximum on the photometric major axis) that is similar to what \cite{2020Neumann} (Fig.14 of their paper) measured for nearby massive barred galaxies (with M$_{\star}$ $>$ 10$^{10}$ M$_{\odot}$) \citep[also predicted in Auriga hydrodynamical cosmological simulations by][]{2017Grand}. The bar's potential can trap stars in quasi-periodic or periodic orbits, based on their initial kinematics or that of the gas they formed from \citep[e.g.,][]{2017Fragkoudi}. \cite{2020Neumann} showed that bars of massive galaxies trap young and intermediate-age stars ($<$ 8 Gyr) in elongated orbits, shaping a V-shape profile, which we could also detect in the [M/H] and [$\alpha$/Fe] maps of this low-mass dE with log(M$_{\star}$/M$_{\odot}$) = 8.9.
\end{itemize}

\subsection{Star formation histories} \label{Results_2}

\begin{figure*}
    \centering
	\includegraphics[scale=0.45]{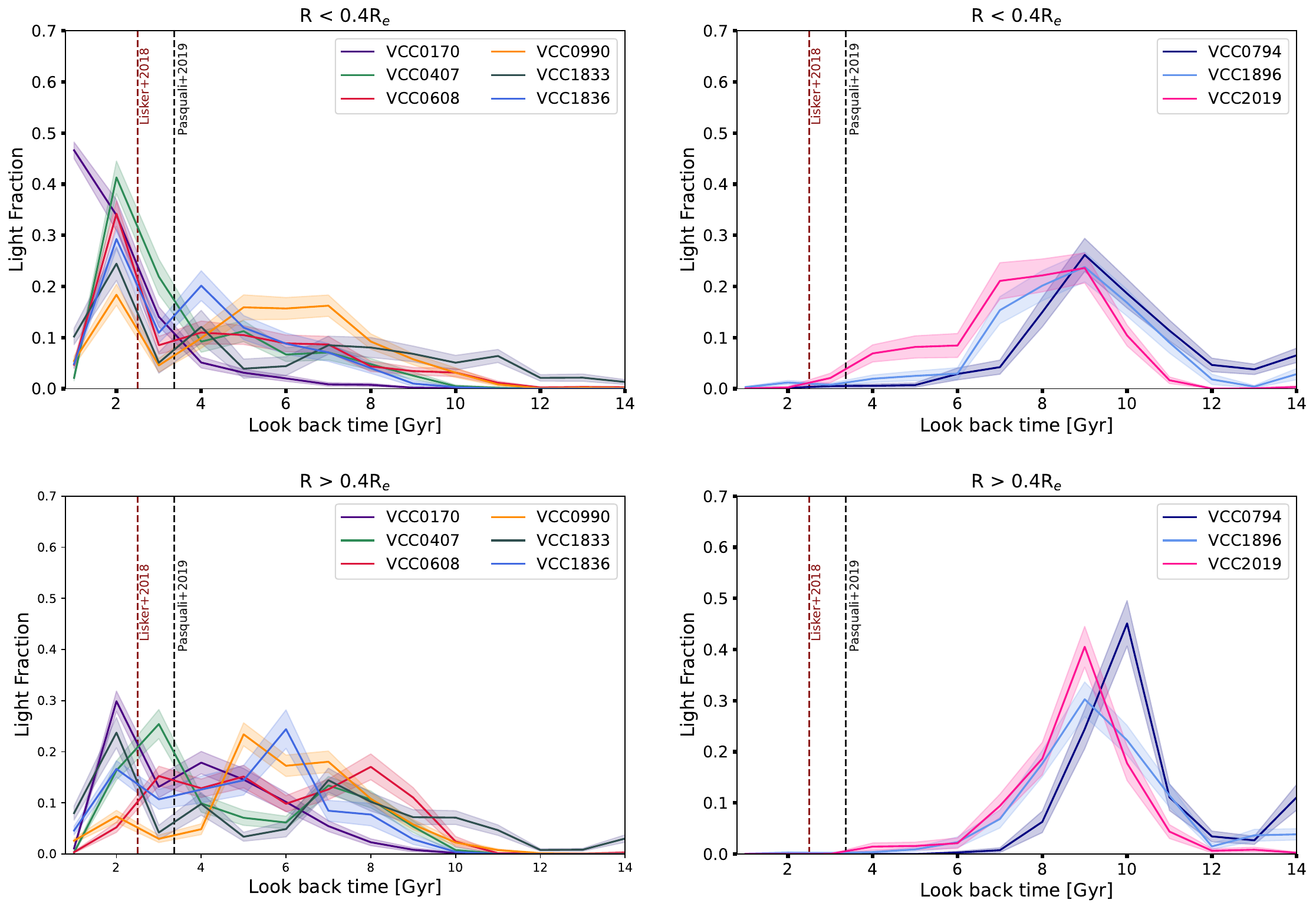}
	\caption{Top row: The stellar light fraction ($\epsilon_{t,Z\rvert\alpha}$) of our sample of dEs as a function of time within an aperture of 0.4$R_{\rm e}$. The SFH profiles of those dEs that exhibit a prominent peak after their accretion onto the Virgo cluster are presented in the left-hand panel, and profiles of those dEs without a similar peak are shown in the right-hand panel. Bottom row: Same as the top row but for outer parts of galaxies with R>0.4R$_{\rm e}$. The red and black dashed vertical lines indicate the accretion time of these nine dEs onto the Virgo cluster as predicted by Lisker et al. (2018) and Pasquali et al. (2019), respectively (c.f., B22). }
	\label{SFH profiles}
\end{figure*}

We construct star formation history (SFH) profiles from results of STARLIGHT full spectrum fitting, as explained in Section \ref{Analysis_2}. Each STARLIGHT run provides a population vector, whose components are the light fractions ($\epsilon_{t,Z\rvert\alpha}$) contributed by each SSP to the best-fitting spectrum. In Fig. \ref{SFH profiles}, we present the stellar light fraction as a function of look-back time for dEs inner and outer regions. Here the stellar light fractions are the sum of all ($\epsilon_{t,Z\rvert\alpha}$) of the base models with the same age but different [$\alpha$/Fe] and [M/H]. These solid lines in Fig. \ref{SFH profiles} are the averages over 100 profiles constructed from MC iterations (explained in Section \ref{Analysis_2}). The shaded area around solid lines are the standard deviation of the derived profiles. The inner and outer profiles of each dE have been normalized by the galaxy's total flux. 

Similar to B22, we divide our sample into two sub-groups based on the presence or absence of a recent peak of star formation during or after their accretion onto Virgo. The SFH profiles of galaxies with recent peaks are shown on the left-hand columns of Fig. \ref{SFH profiles}, while those without any recent prominent peak are presented in the right-hand columns. Panels on the top show the SFHs in the inner regions (i.e., R$<$ 0.4R$_{\rm e}$), while the SFHs in the lower panels belong to the outskirts (i.e., R$>$ 0.4R$_{\rm e}$). Two dashed vertical lines indicate two estimates for the average infall time of these nine dEs (i.e., 2-3 Gyr) based on the work done by \cite{Lisker2018} (the red line) and \cite{2019Pasquali} (the black line).

Our results in Fig. \ref{SFH profiles} show that dEs with a recent peak of star formation (left panels) are on average younger, even in their outskirts, than their counterparts without any recent star formation activity (right panels). As the profiles of the six dEs on the top left panel indicate, the recent phase of star formation (i.e., between 4 to 2 Gyr ago) is mainly confined in the inner part of these galaxies. The stellar population maps of these six dEs confirm the presence of a younger, more metal-rich, and less $\alpha$-enhanced stellar population in their inner regions. Moreover, our results in the lower left-hand panel of Fig. \ref{SFH profiles} indicate that the mass assembly in the outskirts was slow and took place over an extended time scale.

Three dEs without a recent peak of star formation, shown on the right-hand panels of Fig. \ref{SFH profiles}, have completed their mass assembly long before their accretion onto Virgo. Their mass assembly beyond 0.4 R$_{\rm eff}$ was completed more than $\sim$ 6 Gyr ago and on a rather short timescale. The mass assembly in their inner regions was completed $\sim$ 2 Gyr prior to their accretion onto Virgo, and lasted longer. These results are consistent with the presence of metal-poor and $\alpha$-enhanced stellar populations at large effective radii of these three galaxies, shown in maps of Fig. \ref{stellar pops maps1} and \ref{stellar pops maps2}. We note that the SFHs of these three dEs were shorter than that of the remaining six (in the left-hand panel) both in the inner regions and outskirts.

\subsection{Radial gradients of stellar population properties}\label{Results_3}
We quantify the spatial variation of the stellar population properties (and indices) by assuming a linear relation between them and log (R/Re): 
\begin{equation}
    \rm Param(R/R_{e}) = Param_{Re} + \nabla_{param} \times log(R/Re)
\end{equation}
where Param$_{\rm Re}$ represents the fitted stellar population property (or Lick index) at the galaxy effective radius, and $\nabla_{\rm param}$ is the slope of the fitted line (hereafter referred to as gradient). In Appendix \ref{Radial profiles} we present the radial profiles of age, [M/H], and [$\alpha$/Fe] of each dE, along with the fitted linear line. To account for biases due to the effect of seeing, we excluded bins within 1.5FWHM$_{\rm seeing}$ (marked as a grey-shaded area in Fig. \ref{Radial_prof1} and \ref{Radial_prof2}) prior to the fitting of the line. As mentioned in Section. \ref{Results_1} VCC0990 and VCC2019 host a central NSC whose presence can affect the linear fitting. Therefore, for these two dEs, we excluded bins within the NSC radius that we measured in B22. The yellow shaded area in the radial profiles of these two dEs, as shown in Fig. \ref{Radial_prof1} and \ref{Radial_prof2} represents the NSC radius. 

To fit the radial profiles, we employ the publicly available LINmix routine \citep{2007Kelly}. LINmix is a Bayesian method that performs linear regression over data while accounting for measurement errors. The slopes and intercepts reported in Table \ref{gradiets_stellar_pops} for age, [M/H] and [$\alpha$/Fe] are the averages of $\sim$ 1000 MC iterations performed by LINmix. Here we assume the standard deviation of these MC iterations as the error on the fitted slopes and intercepts, and we report them in Table \ref{gradiets_stellar_pops}. The slopes, intercepts and their corresponding errors measured for the spectral indices are reported in Table \ref{gradiets_stellar_popsII} and \ref{gradiets_stellar_popsIII} of Appendix \ref{Grad_indices_appendix}. In order to assess the impact of the intrinsic scatter of stellar population properties, at a given radius, on gradients' estimates,
we have also performed the following test. For each bin inside a galaxy, we selected the nearest (N=7) bins, and computed the
standard deviation of their stellar population properties (i.e., age, [M/H], and [$\alpha$/Fe]). We then re-run LINmix adopting the standard deviation estimates as new error bars on binned data points. This procedure leads to negligible differences (within $\sim$1$\sigma$) with respect to  our initial estimates (in Table \ref{gradiets_stellar_pops}), confirming the robustness of our results.
We also note that our fitting procedure is insensitive to the scatter along the y-axis, as it minimizes the square residuals along the y direction.

\setlength{\tabcolsep}{1.5pt}
\begin{table*}
\caption{\label{gradiets_stellar_pops} Gradients of stellar population properties of the main sample dEs}
\centering
\begin{tabular}{c c c c c c c}
\hline
Object     & log(Age)$_{\rm Re}$ & $\nabla_{\rm age}$ &  [M/H]$_{\rm Re}$ & $\nabla_{\rm [M/H]}$ & [$\alpha$/Fe]$_{\rm Re}$ & $\nabla_{\rm [\alpha/Fe]}$\\
    & [log yr] & [log yr]/[log(R/Re)] & [dex] &   & [dex] & \\
\hline
\hline
VCC0170&  10.16 $\pm$0.08 &  0.99 $\pm$0.12  &  -0.80 $\pm$ 0.04 & -0.25 $\pm$0.06  & 0.36 $\pm$0.05 & 0.12 $\pm$0.09\\[0.1cm]
VCC0407& 9.97 $\pm$ 0.04   &  0.50 $\pm$0.1  &  -0.75 $\pm$0.02  & -0.38 $\pm$0.05  & 0.27 $\pm$0.02  & 0.14 $\pm$0.04\\[0.1cm]
VCC0608& 9.78 $\pm$ 0.06   &  0.11 $\pm$0.11  &  -0.71 $\pm$0.02  &-0.21 $\pm$0.04   & 0.42 $\pm$0.02 & 0.26 $\pm$0.03\\[0.1cm]
VCC0794& 9.98 $\pm$ 0.02  &  0.00 $\pm$0.02  &  -0.90 $\pm$0.07  & -0.18 $\pm$0.09  & 0.33 $\pm$0.09 &-0.07 $\pm$0.12\\[0.1cm]
VCC0990&  10.00 $\pm$ 0.01  & -0.01$\pm$0.12 & -0.65 $\pm$ 0.1  & -0.29 $\pm$ 0.04    & 0.32 $\pm$ 0.02        & 0.31 $\pm$0.07   \\[0.1cm]
VCC1833& 9.84 $\pm$ 0.02  &  0.64 $\pm$0.09 &  -0.55 $\pm$0.01  &-0.51 $\pm$0.05   & 0.29 $\pm$0.01 & 0.24 $\pm$0.05\\[0.1cm]
VCC1836& 9.96 $\pm$ 0.07  &  0.12 $\pm$0.10  &  -0.90 $\pm$0.05  &-0.19 $\pm$0.07   & 0.34 $\pm$0.06 & 0.11 $\pm$0.09\\[0.1cm]
VCC1896& 10.00 $\pm$ 0.01 & -0.01 $\pm$0.03  &  -0.65 $\pm$0.01  &-0.10$\pm$0.05    & 0.30 $\pm$0.03 & 0.07 $\pm$0.08\\[0.1cm]
VCC2019& 10.00 $\pm$ 0.01 & -0.02$\pm$0.03   &  -0.72 $\pm$0.02  &-0.19$\pm$0.04    & 0.47 $\pm$0.05 & 0.36 $\pm$0.12\\[0.1cm]
\hline
\end{tabular}\\
Columns are: Name of the target, log(age) at 1Re, slope of age gradient, [M/H] at 1Re, slope of [M/H] gradient, [$\alpha$/Fe] at 1Re, and slope of [$\alpha$/Fe] gradient. \\
\end{table*}

We detect significant positive age gradients in three dEs of our sample, namely VCC0170, VCC0407 and VCC1833. The rest of the sample shows insignificant age gradients. All nine dEs investigated in this study show significant negative metallicity gradients (within the range of  -0.38 < $\nabla_{\rm [M/H]}$ < -0.10), indicating the presence of more metal-rich stellar populations in their central regions. VCC1836 with $\nabla_{\rm [M/H]}$ = -0.38 and VCC1896 with $\nabla_{\rm [M/H]}$= -0.10 have the largest and the smallest metallicity gradients, respectively. These results are consistent with other studies of Virgo and Fornax dEs \cite[e.g.,][]{2008Chilingarian,2009Koleva,2011Koleva,2017Sybilska}. All of targeted dEs (except VCC0794) show positive [$\alpha$/Fe] gradients in the range of -0.07 < $\nabla_{\rm [\alpha/Fe]}$ < 0.36. The $\nabla_{\rm [\alpha/Fe]}$ is the largest in VCC2019 and smallest (near zero) in VCC0794. Positive $\nabla_{\rm [\alpha/Fe]}$ and negative $\nabla_{\rm [M/H]}$ indicates that star formation in our sampled dEs was more extended (and efficient) in the inner regions, relative to the outskirts.

\subsection{Comparison with dEs in the literature} \label{Results_4}
\begin{figure*}
    \centering
	\includegraphics[scale=0.60]{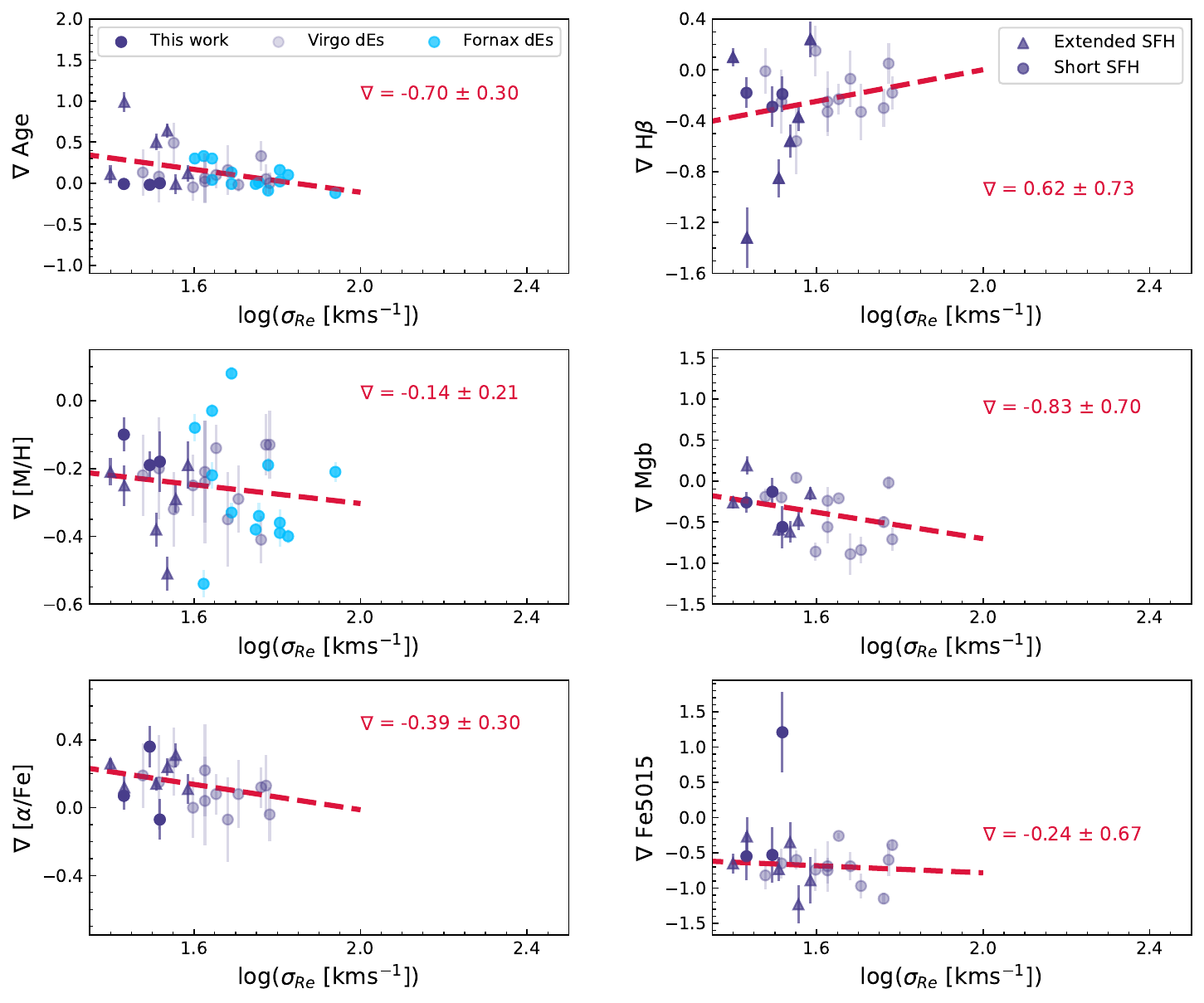}
	\caption{\textit{Left columns:} From top to bottom, $\nabla_{\rm age}$, $\nabla_{\rm [M/H]}$ and $\nabla_{\rm [\alpha/Fe]}$ of our sample dEs (denoted with dark purple data points), S17's Virgo dEs (denoted with light purple data points) and Fornax dEs from K11 (denoted with light blue data points) as a function of log($\sigma_{\rm Re}$). Main sample dEs with short and extended SFHs are marked with circles and triangles, respectively. Fitted linear relations on each sample are shown with red dashed lines. Slopes and their corresponding errors are reported on each panel. \textit{Right columns:} From top to bottom, $\nabla$H$\beta$, $\nabla_{\rm Mgb}$, and $\nabla$Fe5015 are presented for our sample dEs and S17's Virgo dEs. }
	\label{dEs_comparison_gradients}
\end{figure*}

On the left-hand panels of Fig. \ref{dEs_comparison_gradients} and from top to bottom, we plot gradients of age, [M/H] and [$\alpha$/Fe] with respect to logarithmic stellar velocity dispersion measured at the galaxy's R$_{\rm e}$ (log($\sigma_{\rm Re}$)). This figure compares gradients for dEs in our main sample (denoted with dark purple) with those dEs from Virgo and Fornax clusters in light purple and blue, respectively. Main sample dEs with extended and short SFHs, according to Section \ref{Results_2} are marked with circles and triangles, respectively. Note that gradients of [$\alpha$/Fe], as well as gradients of spectral indices (see below) are not available for Fornax dwarf dEs. We choose to plot stellar velocity dispersion rather than stellar mass as it is directly measured from observations, and is less affected by the assumptions made to derive the stellar mass. In each panel, the dashed line represents the linear fit we performed with the LINmix routine (see Section \ref{Results_2}) to evaluate the dependence of each parameter on log($\sigma_{\rm Re}$). Slopes of these lines are reported in the corresponding panels. 

The majority of dEs show flat $\nabla_{\rm Age}$ which is similar to what is reported for dEs \citep[][]{2017Sybilska} and more massive ETGs \citep[e.g.,][]{2007Annibali}. Our results on the top left panel of Fig. \ref{dEs_comparison_gradients} show a negative correlation between $\nabla_{\rm Age}$ and log($\sigma_{\rm Re}$) indicating that for low $\sigma_{Re}$ dEs, stellar populations are younger in their central regions relative to the outskirts. This trend is mainly driven by few dEs in our main sample, like VCC0170 and VCC1833, that have been forming stars (or are still forming stars) until very recently.

Dwarf ellipticals of all the three samples investigated here show negative $\nabla_{\rm [M/H]}$ (down to -0.55, except one), with an average value of -0.2. This result is in good agreement with the work of \cite{1999Kobayashi} and \cite{2008Pipino}, according to which a range of -0.5 $<$ $\nabla_{\rm [M/H]}$ $<$ -0.2 for ETGs is expected (see also \cite{2007Annibali}).
We could not find any statistically meaningful correlation between $\nabla_{\rm [M/H]}$ and log($\sigma_{\rm Re}$) for the present sample of dwarf galaxies. Note that the scatter of $\nabla_{\rm [M/H]}$ is significant, in particular for Fornax dEs. 

 The different scatter  might be due to the fact that observations have been done with different instruments for the three samples of dEs. Fornax dEs have been observed by K11 with FORS2 long-slit spectroscopy (with R = 1666 at $\lambda$ = 5000  \AA), while both samples of Virgo dEs have been observed with IFUs (MUSE with R = 1992 and SAURON with R = 1785, at $\lambda$ = 5000  \AA), providing better spectral resolution and more homogeneous spatial coverage. Being the K11 sample dominated by nucleated dEs \citep[see][]{2022Su} we can expect their spectra as well as their stellar populations to be affected by these nuclear star clusters, and possibly by other substructures (e.g., bar, disk) when present, thus explaining their larger scatter in stellar properties (see K11).
 
We marginally detect a weak negative correlation between the $\nabla_{\rm [\alpha/Fe]}$ and log($\sigma_{\rm Re}$) in the sense that toward more massive dEs, $\nabla_{\rm [\alpha/Fe]}$ decreases. This trend is though significant only at a 1$\sigma$ level, and thus in general is consistent with the work of \cite{2017Sybilska} who could not find any correlation between $\nabla_{\rm [\alpha/Fe]}$ and $\sigma_{\rm Re}$ of low-mass ETGs. 

We also note some differences in the radial gradients of the main sample dEs depending on their SFHs. Those dEs with extended SFHs have steeper $\nabla_{\rm age}$ with an average value of 0.39 $\pm$ 0.14, compared to the remaining three which show flat radial age profiles with an average slope of -0.01 $\pm$ 0.00. Similarly, dEs with long SFHs have steeper $\nabla_{\rm [M/H]}$ (with an average value of -0.30 $\pm$ 0.04) compared to the other three galaxies whose SF is not extended (with an average slope of -0.16 $\pm$ 0.02). However, there is no significant difference between these two sub-samples for what concerns their radial [$\alpha$/Fe] profiles as dEs with short and extended SFHs show an average $\nabla_{\rm [\alpha/Fe]}$ of 0.12$\pm$0.10 and 0.20 $\pm$ 0.03, respectively. 

On the right panel of Fig. \ref{dEs_comparison_gradients} and from top to bottom, we show the gradients of H$\beta$, Mgb and Fe5015 with respect to log($\sigma_{\rm Re}$). We do not detect any meaningful correlation between $\nabla_{\rm H\beta}$ or $\nabla_{\rm Fe5015}$ and log($\sigma_{\rm Re}$). We note that similar to $\nabla_{\rm age}$ and $\nabla_{\rm [\alpha/Fe]}$, $\nabla_{\rm Mgb}$ also shows a negative correlation with log($\sigma_{\rm Re}$), although only at 1$\sigma$ level. We note that \cite{2006Ogando} also found that $\nabla_{\rm Mg2}$ steepens with increasing stellar velocity dispersion in early-type galaxies with log($\sigma_{\rm Re}$) $>$ 2 \citep[see also][]{1993Carollo}.

\subsection{Comparison with ETGs in the literature} \label{Results_5}

\begin{figure}
    \centering
        \includegraphics[scale=0.7]{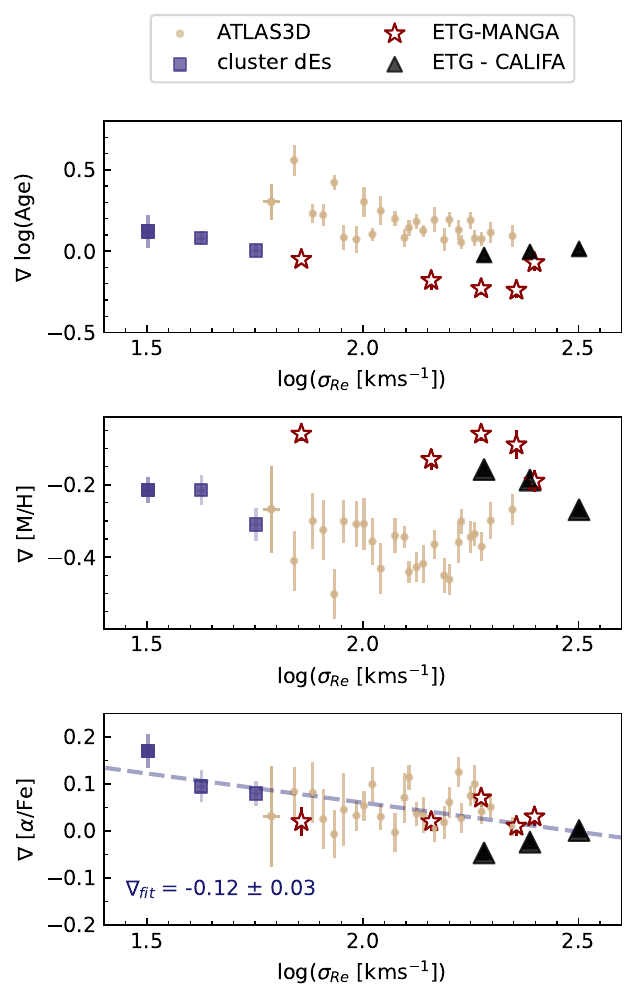}
       \caption{Comparison of the gradients of stellar population properties for cluster dEs (denoted in purple, references in Section \ref{Data2}), ATLAS3D ETGs (denoted in khaki, \protect\cite{2020A&A...635A.129K}), MANGA ETGs (denoted in red, \protect\cite{2021Parikh}) and very massive CALIFA ETGs (denoted in black, \protect\cite{2018MNRAS.475.3700M}). From top to bottom, we compare $\nabla_{\rm age}$, $\nabla_{\rm [M/H]}$, and $\nabla_{\rm [\alpha/Fe]}$, respectively. Note that the MANGA ETG results are derived using binned spectra of galaxies in five stellar mass bins (see Section \ref{Data2}). Data point of ATLAS3D and dEs represents plot median values for different log($\sigma_{\rm Re}$) bins, each including 10 galaxies. The error bars are standard deviation of the median. The dashed line in lower panel traces the linear correlation between $\nabla_{\rm [\alpha/Fe]}$ and log($\sigma_{\rm Re}$). Slope of the fitted line is reported in the same panel.}
            \label{all_comparison_gradients}
\end{figure}
 
The three panels of Fig. \ref{all_comparison_gradients} show how the gradients of different stellar population properties distribute over a wide range of log($\sigma_{\rm Re}$). To this aim, in Fig. \ref{all_comparison_gradients}, we compare dEs of Fig. \ref{dEs_comparison_gradients} (i.e., our sample, plus the comparison samples of the Fornax and Virgo clusters) with massive ETGs. Here, we binned dEs (in dark blue) and ATLAS3D ETGs (in khaki) in log($\sigma_{\rm Re}$) so that each plotted bin includes 10 galaxies. The error bars show the error on the median. Data points indicating the MaNGA (in dark red) and the massive CALIFA ETGs (in black) are directly imported from their original studies. Note that the MANGA ETG results are derived using stacked spectra of galaxies (i.e., not spectra of individual galaxies) in five different stellar masses. This is a biased approach since younger stellar populations in the central regions of even a few galaxies can over-weight the overall derived radial gradient in the stacked spectrum. Thus, they represent gradients computed over different numbers of galaxies per bin. The comparison can be summarized as follows: 
\begin{itemize}
    \item \textbf{Age:} On average, early-type galaxies show similar age gradients in the range of -0.2 $<$ $\nabla_{\rm Age}$ $<$ 0.4, irrespective of their $\sigma_{\rm Re}$. We do not detect any meaningful trend between  $\nabla_{\rm Age}$ and log($\sigma_{\rm Re}$) of ETGs. At log($\sigma_{\rm Re}$ [km$s^{-1}$]) $\sim$ 1.8, where the low-mass ETGs of the ATLAS3D sample and MANGA ETGs overlap with dEs, we witness a noticeable discrepancy among the galaxy samples. dEs and low-mass MaNGA ETGS have a similar $\nabla_{\rm Age}$ $\approx$ 0, while low-mass ATLAS3D galaxies are significantly offset to a more positive $\nabla_{\rm Age}$ by $\sim$ 0.3. 
    
    \item \textbf{[M/H]:} The discrepancy among the galaxy samples is more substantial for $\nabla_{\rm [M/H]}$. While dEs and ATLAS3D ETGs at the overlapping point log($\sigma_{\rm Re}$ [km$s^{-1}$]) $\sim$ 1.8 are consistent with each other, the low-mass MaNGA ETGs diverge significantly. Measurements for MaNGA ETGs at higher log($\sigma_{\rm Re}$ [kms$^{-1}$] $\sim$ 2.4) show more consistency with $\nabla_{\rm [M/H]}$ reported for massive CALIFA galaxies \citep[see also][]{2005Forbes}. dEs and ATLAS3D ETGs alone seem to trace a positive correlation between $\nabla_{\rm [M/H]}$ and stellar velocity dispersion, up to log($\sigma$ [kms$^{-1}$]) $\sim$ 2.1, in the sense that $\nabla_{\rm [M/H]}$ steepens with increasing log($\sigma$ [km$s^{-1}$]). This trend seems to reverse itself for more massive ETGs. Nonetheless, this correlation is mainly set by ATLAS3D data points. At log($\sigma_{\rm Re}$ [kms$^{-1}$]) $\sim$ 2.25, MaNGA and CALIFA ETGs deviate significantly from ATLAS3D. This prevents us from establishing any robust conclusion on possible trends between $\nabla_{\rm [M/H]}$ and log($\sigma_{\rm Re}$) for ETGs. Leaving this discrepancy aside, all ETG samples show negative metallicities, ranging between -0.05 to -0.3, which is in excellent agreement with \cite{1999Kobayashi} and \cite{2008Pipino}, likely because of deeper potential well in the central regions of ETGs. 
    
    \item \textbf{[$\alpha$/Fe]:} The agreement among the different data sets is significantly better for $\nabla_{\rm [\alpha/Fe]}$. Overall, ETGs show positive $\nabla_{\rm [\alpha/Fe]}$ that decreases with log($\sigma_{\rm Re}$), with a slope of -0.12$\pm$0.03 (as indicated in the lowest panel of Fig. \ref{all_comparison_gradients}). dEs have the steepest $\nabla_{\rm [\alpha/Fe]}$ with an average of 0.13 $\pm$ 0.02, compared to ATLAS3D with an average of 0.05 $\pm$ 0.01 or CALIFA and MANGA with an average of $\approx$ 0.0. The steepest $\nabla_{\rm [\alpha/Fe]}$ is observed in dEs with the lowest $\sigma_{\rm Re}$.

\end{itemize}

Fig. \ref{all_comparison_gradients} highlights significant discrepancies among different ETG samples, which we believe are mainly rooted in their different data sets, adopted stellar population models, and methodologies. For this reason, although we trust the relative trends within each dataset, we warn the reader on drawing incautious conclusions from the comparison of the absolute values in different samples in Fig. \ref{all_comparison_gradients}.

%-------------------------------------------------------
\section{Discussion}\label{Discussion}

\subsection{Comparing with results of B22}\label{Discussion1}
In B22, we investigated the integrated star formation history and the stellar population properties of nine Virgo dEs (the main sample here, Section \ref{Data1}) using their averaged MUSE spectra. We showed that they are, on average, metal-poorer and significantly $\alpha$-enhanced than other equally-massive dEs in the Virgo and Coma clusters. In B22 we also showed that these dEs have experienced different star formation histories: three dEs formed their present-day mass fast and long before their accretion onto this cluster, and the remaining six dEs had an extended star formation phase  with a prominent recent peak (i.e., <4 Gyr) around their accretion onto Virgo (see Section \ref{Results_2}). Below we compare the results of this work with those of B22:

\begin{itemize}
    \item Maps of light-weighted stellar population properties for three dEs (namely VCC0794, VCC1896 and VCC2019) presented in Section \ref{Results_1} indicate the presence of old, metal-poor and $\alpha-$enhanced stars all over these galaxy. Hand in hand with their SFH profiles, shown in Fig. \ref{SFH profiles}, our results support B22’s interpretations that the high [$\alpha$/Fe] and low [M/H] in these three dEs can be due to their rapid star formation and its early cessation which has taken place, long before their accretion onto Virgo and possibly due to to the effect of the parent group.
    \item The remaining six dEs host relatively younger, more metal-rich and less $\alpha$-enhanced stars in their central regions (in Figs. \ref{stellar pops maps1} and \ref{stellar pops maps2}). This corresponds to the enhanced star formation episode that they have experienced mainly in R$<$ 0.4Re, during or after their infall onto Virgo. Our results are in agreement with B22 where we suggested ram pressure during or after the accretion event as the most plausible explanation for their short and recent phase of star formation, measured from the integrated MUSE spectra. As explained by \cite{2001Vollmer}, ram pressure strips gas from infalling galaxies outskirts but at the same time triggers star formation in their inner regions by increasing the gas surface density, similar to what we see for these six dEs. 
\end{itemize}

Regardless of their SFH (being extended or short), our main sample dEs show on average similar levels of $\alpha$-enhancement (i.e., $[\alpha$/Fe] $\approx$ 0.3 dex) even in their inner regions, which is higher than what is usually observed for dEs of similar mass \citep[e.g.,][]{2008Sansom, 2021Gallazzi,2023MNRAS.tmp..923R}. The work carried out here supports the interpretation we presented in B22, that the high [$\alpha$/Fe] of the main sample dEs could be the result of the short phase of star formation they experienced after falling onto Virgo and(or) a more efficient quenching of star formation, at early-times, possibly because of pre-processing in the previous parent group. 

\subsection{Mass assembly in dEs}\label{Discussion2}
As Fig. \ref{dEs_comparison_gradients} shows, dEs are mainly characterized by flat $\nabla_{\rm age}$, negative $\nabla_{\rm [M/H]}$ and positive $\nabla_{\rm [\alpha/Fe]}$. Radial gradients of stellar population properties trace chemical enrichment processes and star formation at different radii \citep[e.g.,][]{1998Martinelli,2006Pipino, 2008Pipino}. Detailed chemical evolution models of \cite{2008Pipino} suggest that in a dissipative collapse, stellar populations in the outskirts of an elliptical galaxy may form faster than those in the central regions. Stars that formed at earlier epochs evolve and pollute ISM with heavier elements. Hence, new generations of stars, that are mainly born in central regions, are expected to be younger, more metal-rich and less $\alpha$-enhanced. The SFH profiles of our main sample dEs (Fig. \ref{SFH profiles}) and their negative $\nabla_{\rm [M/H]}$ are in good agreement with the expectations of a dissipative collapse model. Moreover,
the range of $-0.6<\nabla_{\rm [M/H]}<0.0$, as well as positive $\nabla_{\rm [\alpha/Fe]}$, reported in the present work for dwarf galaxies (see Fig. \ref{dEs_comparison_gradients}) are in agreement with \cite{2008Pipino,2010Pipino}, who theoretically reproduced the negative slope of $\nabla_{\rm [M/H]}$ $\sim$ -0.3 for ETGs which was observed by several studies \citep[e.g.,][]{1993Carollo,1993Davies,2003Mehlert,2007Annibali,2010Rawle,2011Koleva,2020Santucci}.  

For our main sample, we also noted that dEs with more extended SFHs have steeper $\nabla_{\rm age}$ and  $\nabla_{\rm [M/H]}$ than those with shorter SFHs (at the 3$\sigma$ level). However, the difference in their $\nabla_{\rm [\alpha/Fe]}$ is marginal (less than the 1$\sigma$ level). These findings expand on those presented in Section \ref{Discussion1} that dEs with longer SFHs experienced a recent episode of star formation (as opposed to those with short SFHs) which resulted in younger and more metal-rich stellar populations at smaller radii, and thus, steeper age and metallicity gradients. On the other hand, as discussed in B22, our [$\alpha$/Fe] estimates likely reflect the properties of the old stellar component in each galaxy, as the effect of [$\alpha$/Fe] on absorption lines becomes subdominant for  old relative to young populations \citep[e.g.,][]{2013Conroy}. Therefore, the similar $\nabla_{\rm [\alpha/Fe]}$ for galaxies with  different SFHs in our sample may reflect the similar star-formation time-scale for their old stellar population component, likely because of pre-processing in their parent galaxy group, prior to the accretion onto Virgo.

Since radial gradients of stellar population properties carry fossil records of their host galaxy's mass assembly, it might be expected that gradients correlate with the stellar mass and/or environment of galaxies. Studies have shown correlations between gradients of stellar population properties (mainly for [M/H]) and the stellar mass (or $\sigma_{\rm Re}$) of ETGs \citep[e.g.,][]{2011LaBarbera,2018MNRAS.475.3700M,2020Zibetti}, but debates over possible correlations with environment still go on \citep[][]{2007Annibali,2011LaBarbera,2011Koleva,2015Greene,2018Rosa,2017Zheng,2017Goddard,2019Ferreras,2020Santucci}. In this work, we could not detect any possible environmental effects on the radial distribution of stellar population properties in dEs. In particular, we could not find any correlation between radial stellar population gradients of dEs and their infall time or their host halos' mass (i,e, Fornax vs. Virgo). This can (partly) be due to the small sample size investigated here and(or) negligible effect of environment on the formation of radial gradients of stellar population properties in dEs. 

In Section \ref{Results_5} we also compared our results with more massive ETGs. As discussed there, this comparison suffers from considerable discrepancies between different samples, particularly in [M/H] and age, which prevents us from taking any robust conclusion. Nevertheless, we find a significant correlation between $\nabla_{\rm [\alpha/Fe]}$ and log($\sigma_{\rm Re}$) as the  $\nabla_{\rm [\alpha/Fe]}$ flattens toward more massive ETGs. This could mean that compared to more massive ETGs, dEs are able to preserve their star formation activity in their inner regions for more extended timescales and(or) they experience more efficient star formation episodes at lower effective radii \citep[e.g., see][]{2022MNRAS.515.3472S}.

\subsection{Chemical enrichment history of Virgo dEs}\label{Discussion3}

\begin{figure*}
    \centering
    \includegraphics[scale=0.55]{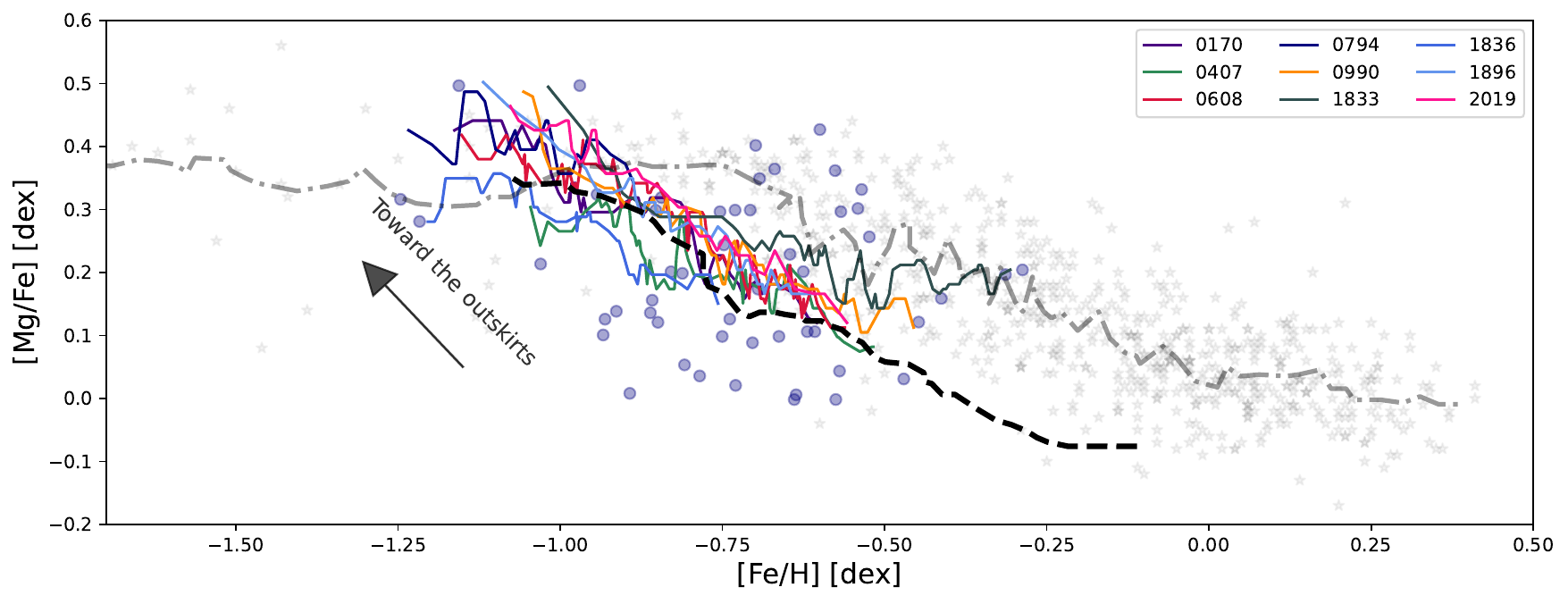}
    \caption{[Mg/Fe] abundance ratio as a function of [Fe/H] for the dEs in the main sample 
    (colour-coded lines), average profile of 20 Virgo dEs of similar stellar mass range \citep[black dashed line,][]{2018Sybilska}, the Large Magellanic Cloud stars \citep[light blue symbols,][]{2008A&A...480..379P}, and Milky Way stars in the Solar neighbourhood \citep[grey symbols,][]{2013Bensby}. The median profile of Milky Way stars is denoted by a gray dashed-dotted line. Relations shown for dEs of the main sample are the running medians of their corresponding bins on the [Mg/Fe]-[Fe/H] space. The dashed black line is the average [Mg/Fe]-[Fe/H] relation over 20 Virgo dEs, as reported by the original study. The arrow on this plot indicates the direction toward larger galactocentric distances. }
	\label{chemical_enrichment_dEs}
\end{figure*}

In the early phases of mass assembly in galaxies, massive stars pollute ISM with heavier elements, such as Mg, C, and N, through SNII explosions. Models predict that, about 1 Gyr after the start of star formation, SNIa gradually dominates the ISM metal enrichment through releasing Fe-like elements \citep{1996Raiteri, 2001Matteucci}. The interplay between these two phases of ISM enrichment, and hence the star formation timescale and efficiency, is often investigated using the [Mg/Fe] vs. [Fe/H] relation. We show this relation for the distribution of individual Milky Way stars in the Solar neighbourhood (hereafter MW) in Fig.\ref{chemical_enrichment_dEs} plotted as gray symbols and adapted from \cite{2013Bensby}. The onset of SNIa manifests as a turnover in $\alpha$ abundances in the [Mg/Fe] vs. [Fe/H] relation. This turnover is known as the “$\alpha$ knee” and as for the MW, it is located at [Fe/H] $\approx$ -0.5. 

The [Mg/Fe] vs. [Fe/H] relations for our nine dEs are also presented in Fig.~\ref{chemical_enrichment_dEs} where we show dEs with color-coded lines. These lines are smoother [Mg/Fe] vs. [Fe/H] relations each constructed as the running median of its corresponding distribution on the [Mg/Fe] vs. [Fe/H] diagram. We also include an average [Mg/Fe]-[Fe/H] relation for Virgo dEs reported by \cite{2018Sybilska} (denoted with a dashed black line and adapted directly from Fig. 4 of their work). The latter relation is an averaged profile over 20 Virgo dEs, with stellar mass range of 9.0 < log (M$_\star$ [M$_\odot$]) < 9.7 and averaged accretion time of > 3 Gyr onto Virgo (see B22). The arrow on this plot indicates the direction toward larger galactocentric distances. Individual stars from the Large Magellanic Cloud (LMC) are also shown in Fig. \ref{chemical_enrichment_dEs} \citep[denoted with light blue symbols, ][]{2008A&A...480..379P}. We point out that the LMC, despite its similar stellar mass to our main sample dEs, is a star forming irregular dwarf galaxy. This might explain the large scatter of observed LMC data points in Fig. \ref{chemical_enrichment_dEs}, with respect to the radial trends for dEs and the MW.

Although, we cannot locate the $\alpha$ knee of dEs in Fig.\ref{chemical_enrichment_dEs} but by comparing the chemical enrichment trend of these low mass galaxies with that of the MW, it is evident that the $\alpha$ knee of Virgo dEs locates at lower [Fe/H] than that of the Galaxy \citep{2018Sybilska}. This is expected as studies have shown that the metallicity at which the $\alpha$ knee locates depends on galaxy stellar mass in the sense that the $\alpha$ knee shifts toward higher [Fe/H] values for more massive galaxies \citep{2011Walker, 2012McConnachie,2014deBoer,2018Sybilska}. In particular, the location of $\alpha$ knee in the [Mg/Fe] vs. [Fe/H] plot can be interpreted in light of the galaxy’s star formation efficiency. Higher star formation efficiency (which is expected in more massive galaxies) increases the metal content of the ISM prior to the onset of SNIa, and shifts the location of $\alpha$ knee to higher metallicities \citep[e.g.,][]{2014deBoer, 2015Walcher}. 

The locations and slopes of the [Mg/Fe] vs. [Fe/H] relations of our nine dEs are  similar to each other, likely due to their narrow stellar mass range (8.9 < log (M$_\star$ [M$_\odot$]) < 9.2, B20). This is consistent with other studies, who showed that galaxies of similar stellar mass occupy a similar zone of the [Mg/Fe]-[Fe/H] space and follow a similar chemical enrichment pattern \citep[e.g.,][]{2014deBoer, 2015Walcher, 2018Sybilska}. In fact, mass-metallicity relations \citep[e.g.,][]{2005Gallazzi} show that galaxies with a stellar mass typical of dEs (i.e., log(M$_{\star}$/M$_{\odot}$) $\sim$ 9.0) have an average [Fe/H] $\sim$ -0.6 dex while MW-like galaxies tend to have higher [Fe/H] of about -0.3 dex. Hence, according to the mass-metallicity relation prediction  of\cite{2005Gallazzi}, the median offset of 0.3 dex between dEs and MW on the [Fe/H] axis in Fig. \ref{chemical_enrichment_dEs} is expected due to their difference in stellar mass. 

Despite the similar location and slope of the relations for our main sample and that of Virgo dEs, our nine dEs have truncated [Mg/Fe] vs. [Fe/H] profiles. This trend remains even if we cut \cite{2018Sybilska}’s sample to match the stellar mass range of our galaxies. The truncated profiles of our main sample dEs indicate that, on average, and particularly in their inner regions, they are metal-poorer and more $\alpha$-enhanced than their counterparts, regardless of their SFHs (i.e., being truncated or extended). 

 The truncated [Mg/Fe] vs. [Fe/H] profiles for our sample of dEs could be the result of pre-processing in their parent group, that made their star-formation efficiency  lower and/or quenched them earlier, compared to Virgo dEs. 
Both effects might explain an higher [Mg/Fe], and a lower [Fe/H] in the galaxies' central regions,
as shown by the tracks in Fig.~\ref{chemical_enrichment_dEs}.
This is also consistent with \cite{2011deLaRosa}, who showed that galaxies attain higher [$\alpha$/Fe] as result of a shorter star formation timescale.
In particular, the above scenario would apply to  the three dEs that were accreted onto Virgo already quenched, forming their stars over a short timescale (see Fig. \ref{SFH profiles}). These three dEs indeed have the shortest profiles in Fig. \ref{chemical_enrichment_dEs}, compared to the rest. On the other hand, the remaining six dEs in our sample  experienced a short starburst at the time of infall, mainly within 0.4R$_{\rm e}$. While some of them (e.g. VCC1833 and VCC0990) have more extended [Mg/Fe] vs. [Fe/H] tracks than the other galaxies in our sample, these tracks are still significantly truncated than those of Virgo dEs from \cite{2018Sybilska}. This is likely because our [Mg/Fe] mostly reflects the property of the galaxies' old stellar component (see above), and the young population (formed at the time of infall) gives only a subdominant contribution to the measured (luminosity-weighted) [Fe/H] \citep{2013Conroy}.

%==================================================================
\section{Conclusion}\label{Conclusion}

This work presents a detailed analysis of stellar population properties over a sample of nine Virgo dEs observed with MUSE. Here, we have derived the age, [M/H] and [$\alpha$/Fe] of the stellar component by comparing observed spectral indices with the SSP model predictions utilizing a four-step fitting routine, developed in \cite{2022Bidaran}. We have presented maps (Fig. \ref{stellar pops maps1} and \ref{stellar pops maps2}) and radial gradients of the stellar population properties of these nine dEs (Section \ref{Results_3}) followed by comparisons with other dwarfs and ETGs in the literature. Additionally, we have derived star formation histories of the main sample within and outside of 0.4R$_{e}$, using full spectrum fitting, to provide a more complete picture of the spatial distribution of stellar population properties. From this work, we conclude that:

\begin{itemize}
	\item Except for VCC0170, VCC0407 and VCC1833, the main sample dEs show flat age radial profiles with negligible fluctuations in value (-0.02 $<$ $\nabla_{\rm Age}$ $<$ 0.12). Among the exceptions here VCC0170 is still actively forming stars in its inner regions and the other two show traces of recent star formation episodes. Moreover, the entire main sample show negative metallicity gradients (in the range of -0.38 $<$ $\nabla_{\rm [M/H]}$ $<$ -0.10) and positive [$\alpha$/Fe] gradients (in the range of -0.07 $<$ $\nabla_{\rm Age}$ $<$ 0.36). Our results indicate the presence of relatively younger, more metal-rich and less $\alpha$-enhanced stellar populations in their inner regions. 

    \item The SFH profiles derived in this work (Fig. \ref{SFH profiles}) reveal that three dEs have assembled their total mass rather fast and long before their accretion onto Virgo. Such short and early bursts of star formation can explain their flat age and shallow [M/H] gradients, as well as the truncated [Mg/Fe] vs. [Fe/H] profiles shown in Fig. \ref{chemical_enrichment_dEs}. The rest of the main sample experienced overall extended star formation histories followed by a short but prominent peak, confined within R < 0.4R$_{e}$, at (or after) their accretion onto Virgo. Short [Mg/Fe] vs. [Fe/H] profiles of these six dEs, mainly set by higher [$\alpha$/Fe] values in their inner regions, trace this short starburst event. 

    \item We found that cluster dEs, irrespective of their mass, show a similar range of $\nabla_{\rm age}$ and $\nabla_{\rm [M/H]}$, with the latter being on average $\sim$ -0.2. These results are consistent with theoretical predictions of \citep{2008Pipino} and together with positive $\nabla_{\rm [\alpha/Fe]}$ suggests that star formation continues over larger timescales in dEs’ inner regions. We noted a possible correlation between $\nabla_{\rm age}$ and the log($\sigma_{\rm Re})$ in the sense that dEs with higher log($\sigma_{\rm Re})$ tend to have flat $\nabla_{\rm age}$. However, these findings should be tested over larger samples of dwarfs. 
    
    \item We found a negative correlation between $\nabla_{\rm [\alpha/Fe]}$ and log($\sigma_{\rm Re})$ of ETGs, in the sense that galaxies with larger log($\sigma_{\rm Re})$ than dEs show flatter $\nabla_{\rm [\alpha/Fe]}$. Since the [$\alpha$/Fe] is believed to trace the duration of star formation, the steep (more positive) $\nabla_{\rm [\alpha/Fe]}$ of dEs may reflect a shorter star-formation in their outskirts than in the inner regions. On the other hand, the flat $\nabla_{\rm [\alpha/Fe]}$ of more massive ETGs may be due to the fact that star formation was very short at all galactocentric distances. Intrinsic properties, such as the galaxy's potential well, star formation efficiency and possibly IMF, as well as environmental effects (e.g., local and large scale environments, and pre-processing) could explain the different  timescales, at different galactocentric distances, between dEs and their more massive counterparts. However, this comparison suffers from discrepancies among different samples, and large, homogeneous, datasets are required to further investigate correlations of stellar population gradients and galaxy mass. 

\end{itemize}

\section*{Acknowledgements}
 We acknowledge financial support from the European Union's Horizon 2020 research and innovation program under the Marie Sklodowska-Curie grant agreement no. 721463 to the SUNDIAL ITN network. J.~F-B  acknowledges support through the RAVET project by the grant PID2019-107427GB-C32 from the Spanish Ministry of Science, Innovation and Universities (MCIU), and through the IAC project TRACES which is partially supported through the state budget and the regional budget of the Consejer\'ia de Econom\'ia, Industria, Comercio y Conocimiento of the Canary Islands Autonomous Community. GvdV acknowledges funding from the European Research Council (ERC) under the European Union's Horizon 2020 research and innovation programme under grant agreement No 724857 (Consolidator Grant ArcheoDyn). DAG acknowledges support from STFC grant ST/T000244/1. This research is partially based on data from the MILES project. 

%%%%%%%%%%%%%%%%%%%%%%%%%%%%%%%%%%%%%%%%%%%%%%%%%%
\section*{Data Availability}
This work is mainly based on observations collected at the European Southern Observatory under P98, ESO programmes 098.B-0619 and 0100.B-057. The raw and reduced data can be downloaded from the ESO Science Archive Facility. Results of our analysis are available upon reasonable request. 



\bibliographystyle{mnras}
\bibliography{example} % if your bibtex file is called example.bib

%%%%%%%%%%%%%%%%%%%%%%%%%%%%%%%%%%%%%%%%%%%%%%%%%%

%%%%%%%%%%%%%%%%% APPENDICES %%%%%%%%%%%%%%%%%%%%%

\appendix

\section{List of Index bands}

In Table \ref{Index banpass} we provide the list of Lick indices used in this study. 
\begin{table*}
\caption{\label{Index banpass} List of indices used in this study}
\centering
\begin{tabular}{c c c c c}
\hline
Index & Blue bandpass [\AA] & Central bandpass [\AA] & Red bandpass [\AA] & Reference \\
\hline
\hline
H$\beta$ & 4827.875-4847.875 & 4847.875-4876.625&4876.625- 4891.625& \citep{1998Trager}\\
H$\beta_{o}$ &4821.175-4838.404& 4839.275-4877.097 & 4897.445- 4915.845& \citep{2009Cervantes}\\
Fe5015 & 4946.500-4977.750 & 4977.750-5054.000& 5054.000- 5065.2500& \citep{1998Trager}\\
Mgb5177 & 5142.625 - 5161.375 & 5160.125-5192.625& 5191.375- 5206.375& \citep{1998Trager} \\
Fe5270 &5233.150-5248.150 & 5245.650-5285.650& 5285.650- 5318.150 &\citep{1998Trager}\\
Fe5335& 5304.625- 5315.875 & 5312.125-5352.125& 5353.375- 5363.375& \citep{1998Trager}\\
\hline
\end{tabular}\\
The columns show: index's name, the corresponding blue pseudo-continuum, main absorption feature, red pseudo-continuum, and the references.\\
\end{table*}
%------------------------------------------------------------------
\section{The radial profiles of stellar population properties} \label{Radial profiles}
In Fig. \ref{Radial_prof1} and \ref{Radial_prof2} we present the radial profiles of age, [M/H] and [$\alpha$/Fe] for each dE of the main sample. These values are measured per bin, as explained in Section \ref{Analysis_1}. Then, assuming a linear relation between each parameter and galactocentric distance, we used LINMIX to fit a line over each profile, as described in Section \ref{Results_3}. Fitted lines are shown with dashed lines in each panel, and their corresponding intercepts and slopes are reported in Table \ref{gradiets_stellar_pops}. 

\begin{figure*}
    \centering
    \includegraphics[scale=0.75]{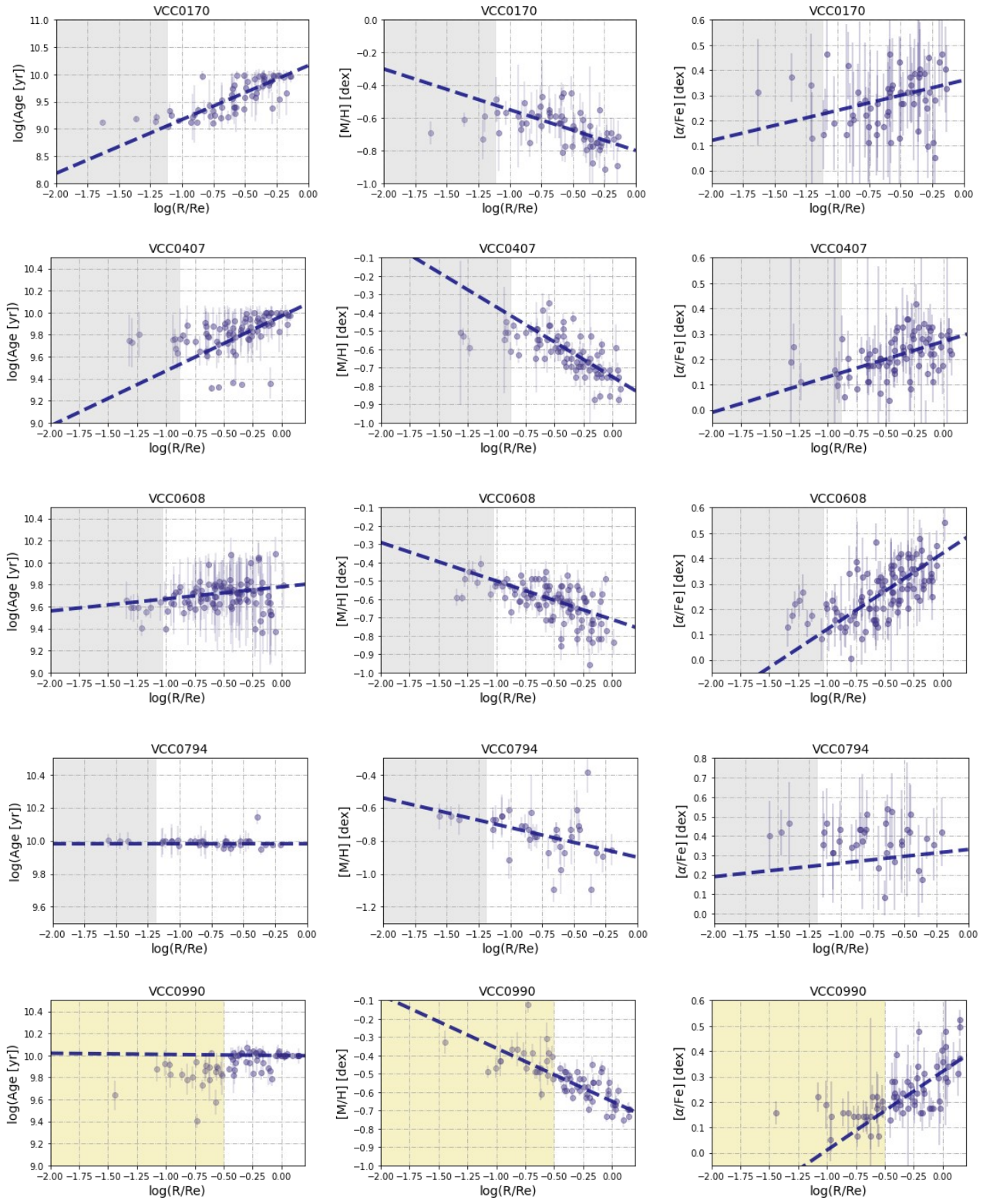}
    \caption{Radial profiles of age (left columns), [M/H] (middle columns), [$alpha$/Fe] (right columns). In each panel, the shaded grey area marks the inner 1.5FWHM$_{\rm seeing}$. Bins within this part are excluded before linear fit. Similarly, the shaded yellow area in panels of VCC0990 and VCC2019 mark the the R$_{\rm NSC}$ and bins within these area are also excluded prior to the linear fit over profiles.  The dashed purple line shows the fitted line over the radial profile and its slope and intercepts are reported in Table \ref{gradiets_stellar_pops}. } 
	\label{Radial_prof1}
\end{figure*}

\begin{figure*}
    \centering
    \includegraphics[scale=0.75]{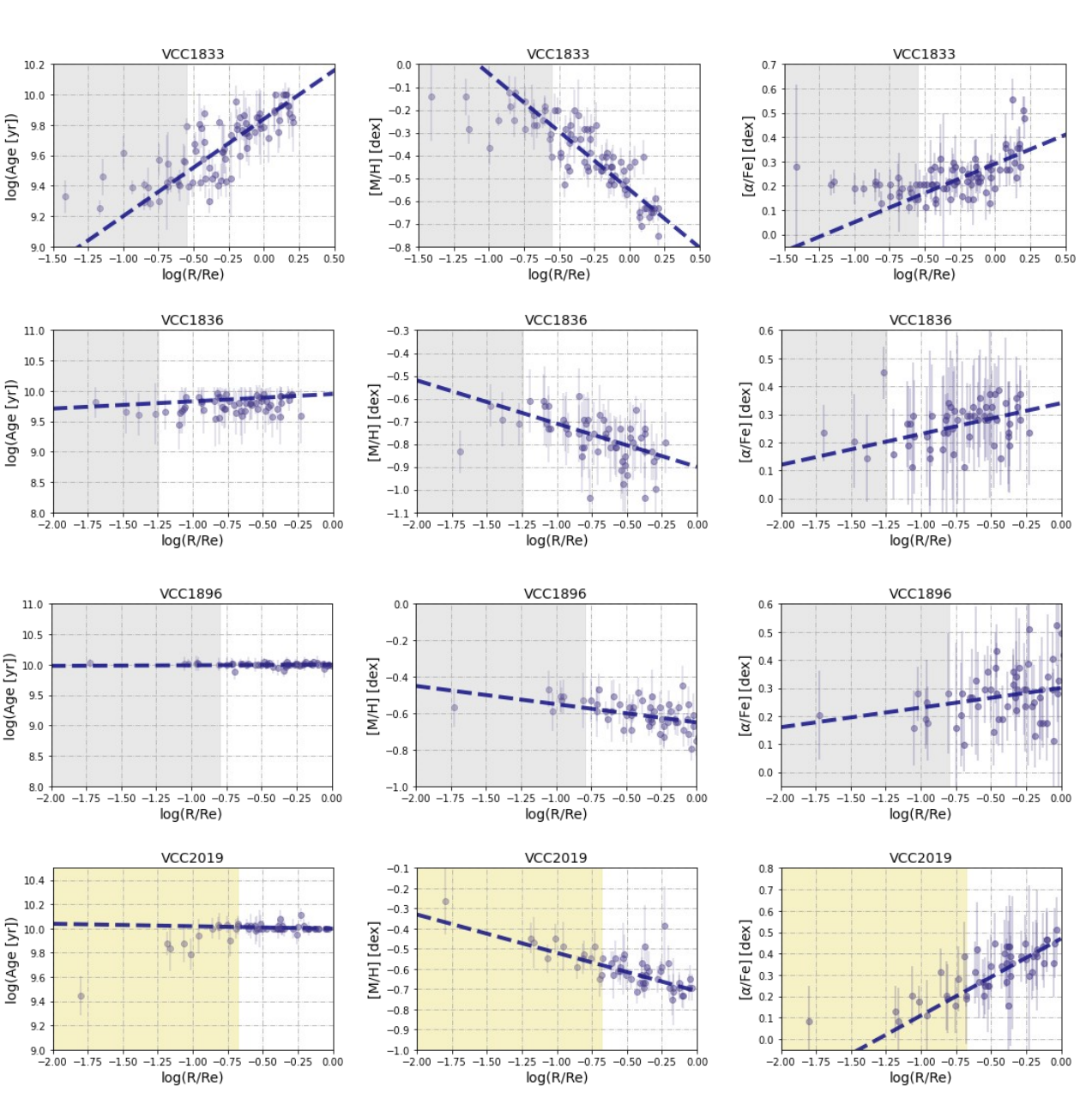}
    \caption{Continued. } 
	\label{Radial_prof2}
\end{figure*}
%---------------------------------------------------------------------------

\section{Gradients of measured spectral indices}\label{Grad_indices_appendix}

In Table \ref{gradiets_stellar_popsII} and \ref{gradiets_stellar_popsIII} we report gradients (i.e., slopes and intercepts) of indices radial profiles for each dE of the main sample. Similar to what is described in Section \ref{Results_3}, we have measured slopes and intercepts of indices radial profiles by employing the public LINMIX routine and by accounting for the measured errors on each index. Errors on the intercepts and slopes reported in this table are the standard deviation of 1000 MC iterations performed by LINMIX as part of its fitting procedure.

\setlength{\tabcolsep}{4.0 pt}
\begin{table*}
\caption{\label{gradiets_stellar_popsII} Gradients of measured indices for the main sample dEs}
\centering
\footnotesize
\begin{tabular}{c c c c c c c}
\hline
Object    & H$\beta_{\rm Re}$    & $\nabla$H$\beta$    & H$\beta_{0,\rm Re}$& $\nabla$H$\beta_{0}$    & Mgb$_{\rm Re}$ &$\nabla_{\rm Mgb}$  \\
    & [\AA]  & [\AA]/[log(R/Re)] & [\AA] &  [\AA]/[log(R/Re)]  & [\AA] & [\AA]/[log(R/Re)]  \\
\hline
\hline
VCC0170 & 2.30  $\pm$ 0.13 & -1.32 $\pm$ 0.24 & 2.86  $\pm$ 0.17 & -2.09 $\pm$ 0.30 & 1.80 $\pm$ 0.06 & 0.19 $\pm$ 0.11 \\[0.1cm]
VCC0407 & 2.20  $\pm$ 0.06 & -0.85 $\pm$ 0.15   & 3.15 $\pm$ 0.07 & -0.97 $\pm$ 0.18 & 1.61 $\pm$ 0.03 & -0.59 $\pm$ 0.08 \\[0.1cm]
VCC0608 & 2.80  $\pm$ 0.03 & 0.10 $\pm$ 0.07   & 3.82 $\pm$ 0.05 & 0.18 $\pm$ 0.09  & 1.73 $\pm$ 0.04 & -0.26 $\pm$ 0.08 \\[0.1cm]
VCC0794 & 1.92  $\pm$ 0.10 & -0.19 $\pm$ 0.14  & 2.52 $\pm$ 0.11 & -0.38 $\pm$ 0.17 & 1.31 $\pm$ 0.18 & -0.56 $\pm$ 0.26 \\[0.1cm]
VCC0990 &  2.12 $\pm$ 0.03 & -0.37 $\pm$ 0.11    & 2.97 $\pm$ 0.05   & -0.80$\pm$0.19  & 2.04   $\pm$ 0.02  &  -0.48 $\pm$ 0.12     \\[0.1cm]
VCC1833 &  2.48 $\pm$ 0.02 & -0.56 $\pm$ 0.13   & 3.45 $\pm$ 0.02 & -0.93 $\pm$ 0.15  & 2.15 $\pm$ 0.02 & -0.62 $\pm$ 0.13 \\[0.1cm]
VCC1836 &  3.01 $\pm$ 0.10 & 0.24 $\pm$ 0.14   & 3.59 $\pm$ 0.10 & -0.15 $\pm$ 0.15 & 1.43 $\pm$ 0.06 & -0.15 $\pm$ 0.08 \\[0.1cm]
VCC1896 &  2.09 $\pm$ 0.04 & -0.18 $\pm$  0.12 & 2.86 $\pm$ 0.05 & -0.52 $\pm$ 0.16 & 2.03 $\pm$ 0.04 & -0.26 $\pm$ 0.13 \\[0.1cm]
VCC2019 &  2.14 $\pm$ 0.07 & -0.29 $\pm$ 0.16  & 2.23 $\pm$ 0.12  & -1.36 $\pm$ 0.31 & 2.07 $\pm$ 0.07 & -0.13 $\pm$ 0.17 \\[0.1cm]
\hline
\end{tabular}\\
Columns are: Name of the target, H$\beta$ at 1Re, the slope of H$\beta$ gradient, H$\beta_{0}$ at 1Re, the slope of H$\beta_{0}$ gradient, Mgb at 1Re, the slope of Mgb gradient.\\
\end{table*}

\setlength{\tabcolsep}{4.0pt}
\begin{table*}
\caption{\label{gradiets_stellar_popsIII} Gradients of measured indices for the main sample dEs}
\centering
\footnotesize
\begin{tabular}{c c c c c c c}
\hline
Object    &  Fe5015$_{\rm Re}$ & $\nabla$Fe5015   &  Fe5270$_{\rm Re}$  & $\nabla$Fe5270  & Fe5335$_{\rm Re}$ &  $\nabla$Fe5335 \\
          & [\AA]              & [\AA]/[log(R/Re)] & [\AA]             &  [\AA]/[log(R/Re)]  & [\AA]          & [\AA]/[log(R/Re)]  \\
\hline
\hline
VCC0170 & 2.70 $\pm$ 0.16  & -0.27  $\pm$ 0.28 & 1.78 $\pm$ 0.07 & 0.04  $\pm$ 0.13 & 1.66 $\pm$ 0.11 & 0.13  $\pm$ 0.20\\
VCC0407 & 3.41 $\pm$ 0.06  & -0.73 $\pm$ 0.17 & 1.89 $\pm$ 0.03 & -0.38 $\pm$ 0.08 & 1.54 $\pm$ 0.03 & -0.48 $\pm$ 0.08\\
VCC0608 &  3.18 $\pm$ 0.07  & -0.65 $\pm$ 0.14 & 1.66 $\pm$ 0.04 & -0.46 $\pm$ 0.08 & 1.34 $\pm$ 0.05 & -0.71 $\pm$ 0.10\\
VCC0794 &  1.63 $\pm$ 0.40  & -1.21 $\pm$ 0.57 & 1.34 $\pm$ 0.15 & -0.46 $\pm$ 0.21 & 1.28 $\pm$ 0.19 & -0.22 $\pm$ 0.25\\
VCC0990 &  3.50  $\pm$ 0.07 & -1.23 $\pm$ 0.27 & 1.91 $\pm$ 0.03 & -0.75 $\pm$ 0.13 & 1.78 $\pm$  0.04& -0.38 $\pm$ 0.14\\
VCC1833 & 3.97 $\pm$ 0.05  & -0.35 $\pm$ 0.28 & 2.06 $\pm$ 0.02 & -0.75 $\pm$ 0.15 & 1.74 $\pm$ 0.03 & -0.60 $\pm$ 0.18\\
VCC1836 & 2.37 $\pm$ 0.23  & -0.89 $\pm$ 0.33 & 1.49 $\pm$ 0.09 & -0.31 $\pm$ 0.14 & 1.20 $\pm$ 0.11 & -0.48 $\pm$ 0.19\\
VCC1896 & 3.24 $\pm$  0.10 & -0.55 $\pm$ 0.34 & 1.85 $\pm$ 0.05 & -0.28 $\pm$ 0.16 & 2.01 $\pm$ 0.07 & 0.29 $\pm$ 0.19 \\
VCC2019 & 3.23 $\pm$ 0.15  & -0.53 $\pm$ 0.39 & 1.70 $\pm$ 0.06 & -0.71 $\pm$ 0.15 & 1.53 $\pm$0.13  & -0.51 $\pm$ 0.30 \\
\hline
\end{tabular}\\
Columns are: Name of the target, Fe5015 at 1Re, the slope of Fe5015 gradient, Fe5270 at 1Re, slope of Fe5270 gradient, Fe5335 at 1Re, and slope of Fe5335 gradient. \\
\end{table*}
%-------------------------------------------------------------------------

\bsp	% typesetting comment
\label{lastpage}
\end{document}